\documentclass[a4paper,11pt]{article}

\usepackage{jcappub} 
\setkeys{Gin}{draft=false}

\usepackage[utf8]{inputenc}
\usepackage{amssymb}
\usepackage{cleveref}
\usepackage{subcaption}
\usepackage{microtype}

\title{From Cosmic Bounces to Primordial Black Holes: The Role of Primordial Magnetic Fields }

\author[a]{Ajeet Singhal,}
\author[a]{and Shreya Banerjee}

\affiliation[a]{Department of Physics, Indian Institute of Technology (Indian School of Mines),\\ Dhanbad, 826004, India}

\emailAdd{23ms0010@iitism.ac.in}
\emailAdd{shreya@iitism.ac.in}

\abstract{
We investigate primordial black hole (PBH) formation from primordial magnetic fields generated during a non-singular ekpyrotic bounce. A time-dependent coupling between the electromagnetic field and a scalar background breaks conformal invariance and amplifies gauge field fluctuations during contraction, generating magnetic fields whose anisotropic stress dominates during radiation domination. By retaining finite-wavenumber cutoffs, we compute the ultraviolet-dominated curvature spectrum, evaluate density-perturbation variance, and determine PBH abundances using both monochromatic and critical-collapse prescriptions. For benchmark parameters, the critical collapse mass spectrum peaks at $M\simeq 2.0\times10^{-3}M_\odot$ , with a total PBH dark matter fraction of $f_{PBH}\approx 10^{-5}$, forming a subdominant subsolar PBH population. Enforcing the cosmological overclosure bound ($f_{\mathrm{PBH}} \le 1$) constrains the primary background parameters, namely the end-of-contraction Hubble scale ($H_{\mathrm{end}} \le 1.6 \times 10^3~\mathrm{GeV}$) and the contraction exponent ($q \le 0.247$), establishing a physical boundary consistent with the electromagnetic backreaction limit. Shifting to asteroid-mass PBHs or a larger dark matter fraction requires modifying the ultraviolet cutoff and contraction duration. This analysis highlights the essential role of finite-wavenumber cutoffs and assumed-collapse prescriptions in linking bounce magnetogenesis to the predicted PBH abundance.
}

\begin{document}
\maketitle
\flushbottom
% Start Page Numbering from Introduction
\section{Introduction}
\label{sec:introduction}
The origin of primordial magnetic fields (PMFs) remains an important open problem in cosmology. Observations of magnetic fields in galaxies and galaxy clusters, together with constraints on intergalactic magnetic fields \cite{RevModPhys.74.775,Grasso_2001,PhysRevD.102.123526,Durrer_2013}, motivate investigations of their possible generation in the early Universe. However, standard Maxwell electrodynamics is conformally invariant in conformally flat Friedmann Lema\^{\i}tre Robertson Walker (FLRW) spacetime, preventing the amplification of electromagnetic vacuum fluctuations solely through the expansion of the Universe. A possible mechanism for breaking this conformal invariance is to introduce a time-dependent coupling $f(\phi)$ between scalar field $\phi$ and the electromagnetic sector through the Lagrangian density $-\frac{1}{4}f^2(\phi)F_{\mu\nu}F^{\mu\nu}$ \cite{1992ApJ...391L...1R,Bamba_2007,Subramanian_2016}. The time dependence of $f(\phi)$ breaks conformal symmetry, enabling quantum electromagnetic fluctuations to grow during early-Universe evolution \cite{Martin:2007ue}.

In this work, we investigate this magnetogenesis mechanism within a non-singular bouncing cosmology \cite{Brandenberger_2017,Ijjas_2018}, where an ekpyrotic contracting phase replaces inflation before transitioning smoothly into a radiation-dominated era. During contraction, the scale factor evolves as $a(\eta) \propto (\tilde{\eta}_{\mathrm{end}} - \eta)^{\tilde{q}}$, decreasing monotonically toward $a_{\mathrm{end}}$. Because the Universe contracts prior to the bounce ($\eta \le \eta_{\mathrm{end}}$), the scale factor satisfies $a(\eta) \ge a_{\mathrm{end}}$ everywhere during the amplification phase, guaranteeing that $f(\phi) = f_{\mathrm{end}} (a/a_{\mathrm{end}})^n \ge 1$ ($n > 0$). This direction of scale-factor evolution inherently avoids the strong-coupling regime ($f \ll 1$) before relaxing to standard electrodynamics ($f \to 1$) post-bounce. Modes exiting the Hubble horizon freeze on super-Hubble scales, yielding a primordial magnetic power spectrum $\mathcal{P}_B(k) = C_B k^{n_B}$ at the end of contraction, where $n_B$ is fixed by the contraction exponent $q$ and coupling index $n$\cite{PhysRevD.94.083524}.

An important consequence of primordial magnetic fields is that their anisotropic stress can source curvature perturbations\cite{PhysRevD.86.023512}. Since the magnetic stress tensor is quadratic in the gauge field, the induced curvature power spectrum $P_{\zeta}(k)$ involves a momentum convolution over pairs of magnetic modes\cite{PhysRevD.86.023512,Kushwaha_2024}. The resulting spectrum depends on the magnetic spectral index, the convolution kernel, and the momentum-integration limits. In particular, when the convolution is dominated by modes near the ultraviolet (UV) cutoff, its scale dependence cannot be inferred reliably from a scale-free approximation alone\cite{PhysRevD.86.023512,Kushwaha_2024}. A consistent calculation must therefore retain the physical wavenumber cutoffs and account for the scale dependence associated with horizon crossing. The resulting curvature spectrum determines the density perturbation variance and provides the input for investigating primordial black holes (PBH) formation during radiation domination\cite{Green:2004wb,Sasaki_2018}.

The central objective of this work is to investigate how primordial magnetic fields generated during a contracting phase source curvature perturbations and influence subsequent PBH formation. Particular attention is given to the role of finite-wavenumber cutoffs in the magnetic convolution and to the propagation of the resulting curvature spectrum into the density variance and PBH mass distribution. The analysis also incorporates the electromagnetic backreaction constraint\cite{PhysRevD.94.083524} and distinguishes between monochromatic and critical collapse prescriptions for PBH abundance\cite{Niemeyer_1998,Byrnes_2018}.

For the benchmark considered here, the model produces a subdominant PBH population in the sub-solar mass range. We also explore how changes in the contraction duration and ultraviolet wavenumber cutoff affect the characteristic PBH mass and abundance, including parameter choices that may permit a large dark matter contribution \cite{Niikura_2019,Carr_2021,Green_2021}.

 Previous studies have investigated primordial magnetogenesis in non-inflationary backgrounds \cite{PhysRevD.94.083524} and the generation of curvature perturbations by magnetic anisotropic stress\cite{Kushwaha_2024}. However, the dependence of the resulting curvature spectrum and PBH abundance on the finite momentum -integration limits requires careful examination, particularly when the convolution is UV-dominated. In this work, we address this issue by evaluating the magnetic convolution with explicit wavenumber cutoffs and tracing its consequences through the density-variance and PBH-abundance calculations. We further compare the monochromatic and critical-collapse prescriptions to clarify how the assumed collapse model affects the predicted PBH mass distribution.

The paper is organized as follows. Section~\ref{sec:bounce_magnetogenesis} introduces the bounce magnetogenesis model, derives the primordial magnetic power spectrum, and discusses the backreaction constraint. Section~\ref{sec:bounce_to_radiation} examines the evolution of gauge field modes across the bounce and into the radiation-dominated era. Section~\ref{sec:numerical_curvature} presents the numerical evaluation of the magnetic sourced curvature spectrum and its analytical scaling, while Section~\ref{sec:density_variance} computes the density perturbation variance. Section~\ref{sec:pbh_framework} establishes the thermal mapping and PBH formation framework. Section~\ref{sec:abundance_formalisms} discusses the monochromatic and critical collapse mass abundance prescriptions. Section~\ref{sec:observational_constraints} presents the PBH mass distributions and their comparison with observational constraints. Finally, Section~\ref{sec:conclusions} summarizes our primary conclusions.

\section{Magnetogenesis in a Bouncing Cosmology}
\label{sec:bounce_magnetogenesis}

\subsection{Background Evolution and Electromagnetic Coupling}
\label{subsec:background_coupling}

We assume a spatially flat Friedmann--Lemaître--Robertson--Walker (FLRW) spacetime described in conformal coordinates by the line element
\begin{equation}
ds^2 = a^2(\eta)\left(-d\eta^2 + d\mathbf{x}^2\right),
\label{eq:metric}
\end{equation}
where $a(\eta)$ is the scale factor and $\eta$ denotes conformal time. The action governing the background bounce dynamics and the non-minimally coupled electrodynamic sector reads \cite{PhysRevD.94.083524}
\begin{equation}
S = \int d^4x \sqrt{-g}\left[ \frac{1}{2}M_{\mathrm{pl}}^2 R + \mathcal{L}_{\mathrm{bounce}} - \frac{1}{4}f^2(\phi) F_{\mu\nu} F^{\mu\nu} \right],
\label{eq:action}
\end{equation}
where $R$ is the Ricci scalar, $M_{\mathrm{pl}} = (8\pi G)^{-1/2} \approx 2.435 \times 10^{18}~\mathrm{GeV}$ is the reduced Planck mass, $\mathcal{L}_{\mathrm{bounce}}$ is the Lagrangian driving the nonsingular bouncing cosmology, $f(\phi)$ is a dimensionless time-dependent coupling function, and $F_{\mu\nu} \equiv \partial_\mu A_\nu - \partial_\nu A_\mu$ is the field strength tensor of the $U(1)$ gauge field.

Breaking Maxwellian conformal invariance ($f(\phi) \neq 1$) allows quantum vacuum fluctuations of the vector field to be amplified during cosmic evolution\cite{PhysRevD.45.1913,Bamba_2007,Subramanian_2016}. The background contracting phase is described by an effective field theory Lagrangian $\mathcal{L}_{\mathrm{bounce}} = -K(\phi,X) - G(\phi,X)\Box\phi$\cite{Cai_2012}, with kinetic variable $X \equiv -\frac{1}{2}\partial_\mu \phi \partial^\mu \phi$. Deep in the contracting epoch ($\phi \ll -1$), the scalar field approaches canonical behavior, yielding a power-law evolution for the scale factor\cite{Brandenberger_2017,PhysRevD.94.083524},
\begin{equation}
a(\eta) = a_{\mathrm{end}} \left( -\frac{\tilde{q}}{a_{\mathrm{end}} H_{\mathrm{end}}} \right)^{-\tilde{q}} (\tilde{\eta}_{\mathrm{end}} - \eta)^{\tilde{q}},
\label{eq:scale_factor}
\end{equation}
where $a_{\mathrm{end}}$ and $H_{\mathrm{end}}$ represent the scale factor and Hubble rate at the end of contraction, respectively. The auxiliary parameters are defined as $\tilde{q} \equiv q/(1-q)$ and $\tilde{\eta}_{\mathrm{end}} \equiv \eta_{\mathrm{end}} - q / \left[(1-q) a_{\mathrm{end}} H_{\mathrm{end}}\right]$, where $q \in (0,1)$ parameterizes the rate of ekpyrotic contraction. The conformal Hubble rate $H(\eta) \equiv a'/a^2$ evaluates to
\begin{equation}
H(\eta) = \frac{\tilde{q}}{\tilde{\eta}_{\mathrm{end}} - \eta} \frac{1}{a(\eta)},
\label{eq:hubble_conformal}
\end{equation}
where primes denote derivatives with respect to conformal time $\eta$.

To avoid ambiguity in the physical background timeline, we explicitly distinguish between two sequential physical events:
\begin{enumerate}
    \item \textbf{End of Contraction ($\eta = \eta_{\mathrm{end}}$):} This hypersurface marks the termination of the ekpyrotic contracting phase. At $\eta_{\mathrm{end}}$, vector mode amplification ceases, the Hubble parameter reaches its maximum contracting magnitude $|H_{\mathrm{end}}|$, and the gauge coupling function terminates at $f(\eta_{\mathrm{end}}) = f_{\mathrm{end}} \sim \mathcal{O}(1)$\cite{PhysRevD.94.083524}.
    \item \textbf{Non-singular Bounce Epoch ($\eta \in [-\Delta\eta_{\mathrm{bounce}}/2, \Delta\eta_{\mathrm{bounce}}/2]$):} The subsequent transition region of finite physical duration $\Delta\eta_{\mathrm{bounce}}$ during which non-canonical higher-derivative terms in $\mathcal{L}_{\mathrm{bounce}}$ dominate, causing the Hubble parameter to pass smoothly through zero ($H = 0$ at the minimal scale factor $a = a_{\mathrm{bounce}}$) and turn around into cosmic expansion\cite{Cai_2012,Brandenberger_2017}.
\end{enumerate}
Because the bounce epoch is brief ($\Delta\eta_{\mathrm{bounce}} \ll |\eta_{\mathrm{end}}|$), super-horizon modes satisfying $k\Delta\eta_{\mathrm{bounce}} \ll 1$ traverse the bounce without mode mixing or spectral distortion\cite{PhysRevD.94.083524,Kushwaha_2024}. Consequently, $a_{\mathrm{end}} \approx a_{\mathrm{bounce}}$ to high accuracy, and initial boundary conditions for post-bounce evolution are consistently fixed at the end of contraction ($\eta = \eta_{\mathrm{end}}$).

During the contracting phase ($\eta \le \eta_{\mathrm{end}}$), the coupling function $f(\eta)$ evolves as a power law of the scale factor\cite{Martin:2007ue,PhysRevD.94.083524},
\begin{equation}
f(\eta) = f_{\mathrm{end}} \left( \frac{a(\eta)}{a_{\mathrm{end}}} \right)^n,
\label{eq:coupling_power_law}
\end{equation}
where $n > 0$ is the coupling exponent. Because the Universe contracts prior to the bounce ($\eta \le \eta_{\mathrm{end}}$), the scale factor decreases monotonically toward $a_{\mathrm{end}}$, enforcing $a(\eta) \ge a_{\mathrm{end}}$ throughout the entire amplification phase. For $n > 0$ and $f_{\mathrm{end}} \sim \mathcal{O}(1)$, this direction of scale-factor evolution guarantees that $\left(a(\eta)/a_{\mathrm{end}}\right)^n \ge 1$, and hence $f(\eta) \ge f_{\mathrm{end}} \sim 1$ everywhere during contraction\cite{PhysRevD.94.083524}. This behavior stands in sharp contrast to inflationary magnetogenesis—where $a(\eta) < a_{\mathrm{end}}$ drives $f \to 0$ in the past—and confirms that ekpyrotic contraction naturally maintains weak gauge coupling ($f \ge 1$) without fine-tuning\cite{PhysRevD.94.083524}. Following the end of contraction, $f(\phi)$ rapidly settles to unity ($f \to 1$), fully restoring standard Maxwell electrodynamics before the radiation-dominated era.

\subsection{Gauge-Field Quantization and Mode Equation}
\label{subsec:gauge_quantization}

Varying action~\eqref{eq:action} with respect to $A_\mu$ yields the modified Maxwell equation\cite{Martin:2007ue,Subramanian_2016},
\begin{equation}
\nabla_\mu \left( f^2 F^{\mu\nu} \right) = 0.
\label{eq:em_eom}
\end{equation}
Fixing the gauge freedom via the Coulomb gauge conditions,
\begin{equation}
A_0 = 0, \qquad \partial_i A^i = 0,
\label{eq:coulomb_gauge}
\end{equation}
The spatial vector potential is quantized in Fourier space as
\begin{equation}
A_i(\eta, \mathbf{x}) = \sum_{\sigma=1,2} \int \frac{d^3k}{(2\pi)^{3/2}} \left[ \epsilon_i^{\,\sigma}(\mathbf{k}) a_{\mathbf{k},\sigma} A_k(\eta) e^{i\mathbf{k}\cdot\mathbf{x}} + \mathrm{h.c.} \right],
\label{eq:A_fourier}
\end{equation}
where $\epsilon_i^{\,\sigma}(\mathbf{k})$ are real transverse polarization vectors obeying $k^i \epsilon_i^{\,\sigma}(\mathbf{k}) = 0$ and $\epsilon_i^{\,\sigma}(\mathbf{k}) \epsilon_i^{\,\sigma'}(\mathbf{k}) = \delta_{\sigma\sigma'}$. The ladder operators satisfy standard commutation relations, $[a_{\mathbf{k},\sigma}, a_{\mathbf{k}',\sigma'}^\dagger] = \delta_{\sigma\sigma'} \delta^{(3)}(\mathbf{k}-\mathbf{k}')$.

Substituting Eq.~\eqref{eq:A_fourier} into Eq.~\eqref{eq:em_eom} gives the differential equation for individual mode functions $A_k(\eta)$ with comoving wavenumber $k \equiv |\mathbf{k}|$:
\begin{equation}
A_k'' + 2\frac{f'}{f} A_k' + k^2 A_k = 0.
\label{eq:mode_equation}
\end{equation}
Canonical quantization enforces the Wronskian normalization condition\cite{Martin:2007ue,Subramanian_2016},
\begin{equation}
A_k(\eta) A_k^{*\prime}(\eta) - A_k^*(\eta) A_k'(\eta) = \frac{i}{f^2}.
\label{eq:normalization}
\end{equation}
To eliminate the friction-like term $2(f'/f)A_k'$, we assume the rescaled auxiliary mode variable
\begin{equation}
v_k(\eta) \equiv f(\eta) A_k(\eta),
\label{eq:vk_def}
\end{equation}
which casts Eq.~\eqref{eq:mode_equation} into Schrödinger-like form:
\begin{equation}
v_k'' + \left( k^2 - \frac{f''}{f} \right) v_k = 0.
\label{eq:vk_eom}
\end{equation}

\subsection{Sub-Hubble and Super-Hubble Solutions}
\label{subsec:mode_solutions}

The behavior of $v_k(\eta)$ depends on the relative competition between the spatial gradient $k^2$ and the effective pump potential $f''/f$:

Deep inside the horizon ($k^2 \gg f''/f$) in the asymptotic past ($\eta \to -\infty$), the effective potential is negligible. Equation~\eqref{eq:vk_eom} reduces to $v_k'' + k^2 v_k \approx 0$, and selecting the positive-frequency state fixes the Bunch--Davies vacuum solution\cite{Birrell:1982ix}:
\begin{equation}
v_k(\eta) = \frac{1}{\sqrt{2k}} e^{-ik\eta}.
\label{eq:bunch_davies}
\end{equation}

On super-Hubble scales ($k^2 \ll f''/f$), spatial gradients are suppressed ($v_k'' - (f''/f)v_k \simeq 0$). Reduction of order yields the general super-horizon solution\cite{Martin:2007ue,PhysRevD.94.083524},
\begin{equation}
v_k(\eta) = C_1(k) f + C_2(k) f \int \frac{d\eta}{f^2}.
\label{eq:super_hubble_gen}
\end{equation}
Using $f(\eta) \propto (\tilde{\eta}_{\mathrm{end}} - \eta)^{n\tilde{q}}$, the integral evaluates to $\int d\eta / f^2 \propto (\tilde{\eta}_{\mathrm{end}} - \eta)^{1 - 2n\tilde{q}}$. Re-expressing in terms of the scale factor $a(\eta) \propto (\tilde{\eta}_{\mathrm{end}} - \eta)^{\tilde{q}}$ gives
\begin{equation}
v_k(\eta) = C_1(k) a^n + C_2(k) a^{-n + \frac{1}{q} - 1}.
\label{eq:super_hubble_scale}
\end{equation}
As $a \to 0$ during contraction, the asymptotic evolution is governed by the smaller exponent. Provided $-1 - n + 1/q < n$, the $C_2$ mode dominates\cite{PhysRevD.94.083524}:
\begin{equation}
v_k(\eta) \simeq C_2(k) a^{-n + \frac{1}{q} - 1}.
\label{eq:growing_mode}
\end{equation}

\subsection{Horizon Crossing and the Primordial Magnetic-Field Spectrum}
\label{subsec:magnetic_spectrum}

The integration constant $C_2(k)$ is fixed by matching the super-horizon growing mode with the Bunch--Davies vacuum state at horizon crossing, $a_k |H_k| = k$\cite{PhysRevD.94.083524,Martin:2007ue}. Equating $|v_k(\eta_k)| \approx 1/\sqrt{2k}$ yields
\begin{equation}
C_2(k) = \frac{1}{\sqrt{2k}} a_k^{n + 1 - \frac{1}{q}},
\label{eq:C2_horizon}
\end{equation}
where $a_k$ is the scale factor at horizon entry. Expressing $a_k$ in terms of parameters at the end of contraction via $a_k = a_{\mathrm{end}} \left( k / (a_{\mathrm{end}} H_{\mathrm{end}}) \right)^{-\tilde{q}}$ gives
\begin{equation}
C_2(k) = \frac{1}{\sqrt{2k}} a_{\mathrm{end}}^{n + 1 - \frac{1}{q}} \left( \frac{a_{\mathrm{end}} H_{\mathrm{end}}}{k} \right)^{\tilde{q} \left( n + 1 - \frac{1}{q} \right)}.
\label{eq:C2_explicit}
\end{equation}
Inserting $C_2(k)$ into Eq.~\eqref{eq:growing_mode} determines the super-horizon mode function:
\begin{equation}
v_k(\eta) = \frac{1}{\sqrt{2k}} \left( \frac{a_{\mathrm{end}} H_{\mathrm{end}}}{k} \right)^{\tilde{q} \left( n + 1 - \frac{1}{q} \right)} \left( \frac{a}{a_{\mathrm{end}}} \right)^{-n - 1 + \frac{1}{q}}.
\label{eq:vk_super}
\end{equation}

Converting back to the physical vector potential $A_k(\eta) = v_k(\eta) / f(\eta)$ via $f(\eta) = f_{\mathrm{end}} (a/a_{\mathrm{end}})^n$ yields
\begin{equation}
A_k(\eta) = \frac{1}{\sqrt{2k} f_{\mathrm{end}}} \left( \frac{a_{\mathrm{end}} H_{\mathrm{end}}}{k} \right)^{\tilde{q} \left( n + 1 - \frac{1}{q} \right)} \left( \frac{a}{a_{\mathrm{end}}} \right)^{-2n - 1 + \frac{1}{q}}.
\label{eq:Ak_super}
\end{equation}
The spatial power spectrum of the generated magnetic field is defined as \cite{Subramanian_2016}
\begin{equation}
\mathcal{P}_B(k,\eta) \equiv \frac{k^5 |A_k|^2}{2\pi^2 a^4}.
\label{eq:PB_def}
\end{equation}
Substituting $A_k(\eta)$ into Eq.~\eqref{eq:PB_def} gives
\begin{equation}
\mathcal{P}_B(k,\eta) = \frac{1}{2\pi^2 f_{\mathrm{end}}^2} \left( \frac{a_{\mathrm{end}} H_{\mathrm{end}}}{k} \right)^{\frac{2q(n+3)-6}{1-q}} \left( \frac{a}{a_{\mathrm{end}}} \right)^{-4n - 6 + \frac{2}{q}} H_{\mathrm{end}}^4.
\label{eq:PB_time}
\end{equation}

Evaluating Eq.~\eqref{eq:PB_time} at the end of contraction ($\eta = \eta_{\mathrm{end}}$, $a = a_{\mathrm{end}}$) produces the final primordial magnetic-field spectrum\cite{PhysRevD.94.083524}:
\begin{equation}
\mathcal{P}_B(k) = C_B k^{n_B},
\label{eq:PB_final}
\end{equation}
where
\begin{equation}
C_B = \frac{1}{2\pi^2 f_{\mathrm{end}}^2} |a_{\mathrm{end}} H_{\mathrm{end}}|^{-n_B} H_{\mathrm{end}}^4, \qquad n_B = -\frac{2q(n+3)-6}{1-q}.
\label{eq:CB_nB_def}
\end{equation}
The spectral index $n_B$ sets the scale dependence of the generated magnetic field, with $n_B = 0$ corresponding to a scale-invariant spectrum.

\subsection{Backreaction Constraint and Parameter Sensitivity}
\label{subsec:backreaction}

To preserve the validity of the background cosmology during ekpyrotic contraction, the energy density of the amplified electromagnetic field $\rho_{\mathrm{EM}}$ must remain strictly subdominant to the background scalar-field energy density $\rho_\phi$. We enforce this requirement via the backreaction condition\cite{Demozzi_2009,Kanno_2009,PhysRevD.94.083524}
\begin{equation}
\rho_{\mathrm{EM}}(\eta) < \rho_\phi(\eta).
\label{eq:backreaction_condition}
\end{equation}

During the contracting phase, the energy densities scale with the scale factor as
\begin{equation}
\rho_{\mathrm{EM}} \propto a^{2 - \frac{6}{q}}, \qquad \rho_\phi \propto a^{-\frac{2}{q}}.
\end{equation}
Because $\rho_{\mathrm{EM}}$ grows faster than $\rho_\phi$ as $a(\eta) \to 0$ for $q \in (0, 1/3)$, the ratio $\rho_{\mathrm{EM}} / \rho_\phi$ reaches its maximum at the end of contraction ($\eta = \eta_{\mathrm{end}}$)\cite{PhysRevD.94.083524}. Satisfying Eq.~\eqref{eq:backreaction_condition} at the transition boundary $\eta_{\mathrm{end}}$,
\begin{equation}
\rho_{\mathrm{EM}}(\eta_{\mathrm{end}}) < 3 H_{\mathrm{end}}^2 M_{\mathrm{pl}}^2,
\end{equation}
guarantees that electromagnetic backreaction remains bounded throughout the entire contracting epoch.

Evaluating the integrated electromagnetic energy density at horizon exit and evolving it to $\eta_{\mathrm{end}}$ yields the upper limit on the end-of-contraction Hubble scale\cite{PhysRevD.94.083524}:
\begin{equation}
\frac{|H_{\mathrm{end}}|}{M_{\mathrm{pl}}} < \frac{\sqrt{2}\pi q}{5(1-q)} e^{-N} \equiv \frac{H_{\mathrm{max}}}{M_{\mathrm{pl}}}.
\label{eq:backreaction_bound}
\end{equation}

\begin{figure}[htbp]
    \centering
    \includegraphics[width=0.75\textwidth]{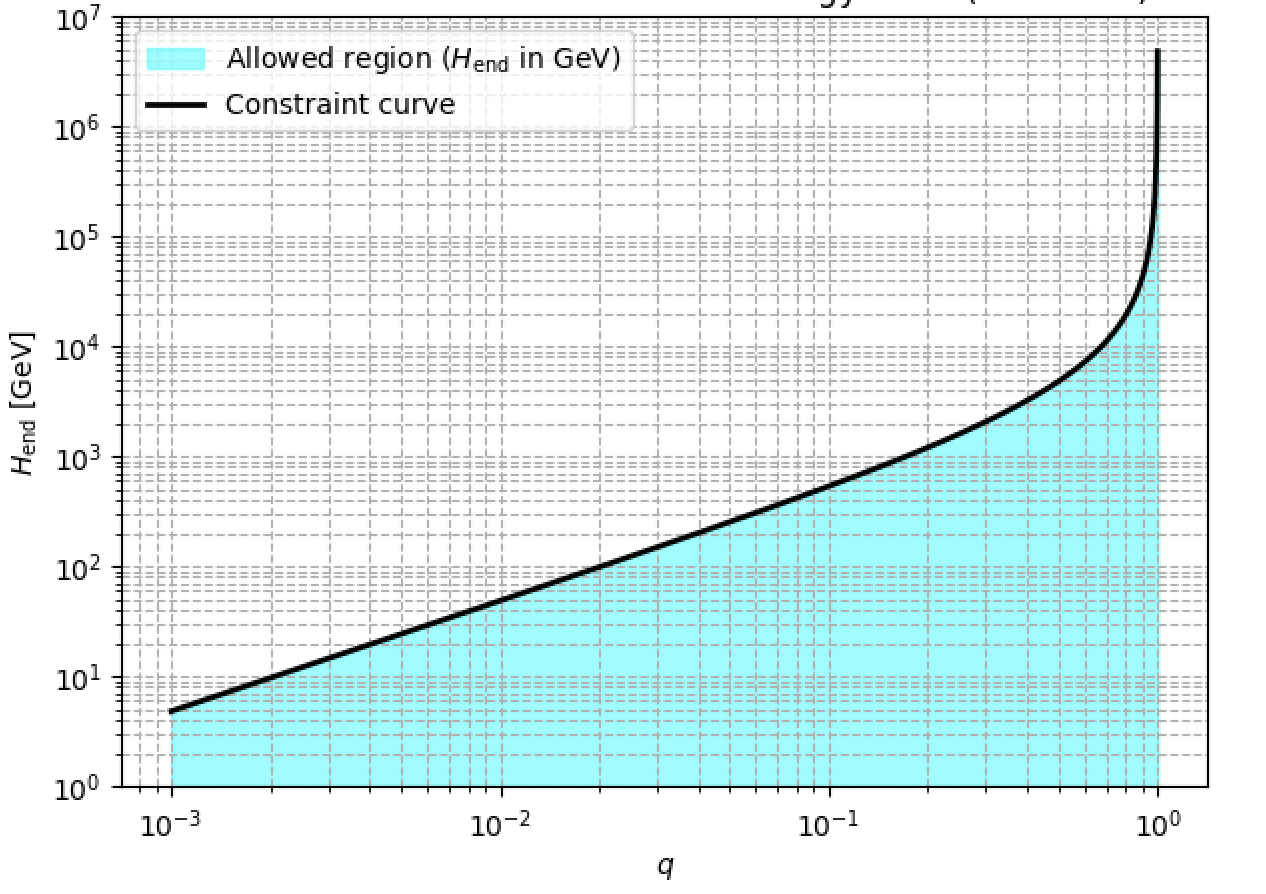}
    \caption{Theoretical upper bound $H_{\mathrm{max}}$ on the end-of-contraction Hubble scale $H_{\mathrm{end}}$ as a function of the background contraction parameter $q$ for $N = 33.74$. The shaded region identifies the parameter domain where the electromagnetic backreaction constraint ($\rho_{\mathrm{EM}} < \rho_\phi$) is satisfied. The solid curve denotes the saturation threshold $H_{\mathrm{max}}(q)$, with the marker indicating the boundary benchmark $H_{\mathrm{end}} = 1493.901~\mathrm{GeV}$ at $q = 0.237$. Slight reductions in $N$ ($N \lesssim 33.70$) shift $H_{\mathrm{max}}$ upward, establishing an explicit safety margin below the saturation line.}
    \label{fig:backreaction_constraint}
\end{figure}

Figure~\ref{fig:backreaction_constraint} shows this theoretical upper bound $H_{\mathrm{max}}$ as a function of $q$ for $N = 33.74$. The shaded region highlights the domain where the electromagnetic energy density remains subdominant ($\rho_{\mathrm{EM}} < \rho_\phi$). For our primary parameter choices $q = 0.237$ and $N = 33.74$, Eq.~\eqref{eq:backreaction_bound} evaluates precisely to
\begin{equation}
H_{\mathrm{max}} = 1493.901~\mathrm{GeV}.
\label{eq:exact_bound_value}
\end{equation}

Setting $H_{\mathrm{end}} = 1493.901~\mathrm{GeV}$ places the background state at the theoretical saturation threshold ($\rho_{\mathrm{EM}} \to \rho_\phi$).

To establish a clear backreaction safety margin, one can slightly adjust the contraction duration $N$. Decreasing $N$ elevates the maximum allowed Hubble scale $H_{\mathrm{max}} \propto e^{-N}$:
\begin{itemize}
    \item At $N = 33.70$ ($q=0.237$), the upper bound increases to $H_{\mathrm{max}} = 1554.87~\mathrm{GeV}$. Retaining $H_{\mathrm{end}} = 1493.90~\mathrm{GeV}$ provides a 3.9 percent physical safety margin ($H_{\mathrm{end}} / H_{\mathrm{max}} \approx 0.961$).
    \item At $N = 33.65$ ($q=0.237$), $H_{\mathrm{max}}$ rises to $1634.59~\mathrm{GeV}$. This gives an 8.6 percent safety margin ($H_{\mathrm{end}} / H_{\mathrm{max}} \approx 0.914$).
\end{itemize}

Throughout this manuscript, we present results for the reference benchmark $H_{\mathrm{end}} = 1493.901~\mathrm{GeV}$ ($N=33.74$) as the maximal saturation boundary for PBH formation in this model. Adjusting $N \approx 33.65\text{--}33.70$ provides a flexible parameter corridor where PBHs form with a robust, explicit backreaction safety margin\cite{PhysRevD.94.083524,Kushwaha_2024}.

\section{Evolution Across the Bounce and into Radiation-Dominated Era}
\label{sec:bounce_to_radiation}

To evaluate the sourcing of primordial curvature perturbations during radiation domination, we trace the post-magnetogenesis evolution of the vector field across the non-singular bouncing cosmology \cite{Cai_2012, Brandenberger_2017, PhysRevD.94.083524}. Because the scale factor $a(\eta)$ drops out of the four-dimensional Maxwell kinetic term due to conformal invariance, the coupling between the scalar background and the gauge field is defined by $z(\eta) \equiv f(\phi)$\cite{Subramanian_2016}. The mode equation for the canonical gauge field $v_k(\eta) \equiv f(\phi) A_k(\eta)$ is given by\cite{Martin:2007ue,PhysRevD.94.083524}
\begin{equation}
v_k'' + \left(k^2 - \frac{f''}{f}\right)v_k = 0,
\label{eq:vk_z_eom}
\end{equation}
where primes denote derivatives with respect to conformal time $\eta$. The physical magnetic energy spectrum is defined as \cite{Subramanian:2015lua,Kushwaha_2024}
\begin{equation}
\mathcal{P}_B(k,\eta) = \frac{k^5 |A_k(\eta)|^2}{2\pi^2 a^4(\eta)}.
\label{eq:PB_physical_def}
\end{equation}

During the ekpyrotic contracting phase, $f(\phi) \propto a^n$ ($n > 0$) drives super-Hubble mode growth\cite{PhysRevD.94.083524}. Following the non-singular bouncing cosmology into the radiation-dominated era, the scalar coupling rapidly settles to $f(\phi) \to 1$, restoring standard electrodynamics such that $z(\eta) \to 1$ and $v_k(\eta) \to A_k(\eta)$\cite{Martin:2007ue,Subramanian:2015lua}.

With $f(\phi) \to 1$, the pump potential vanishes identically in the radiation-dominated era,
\begin{equation}
\frac{z''}{z} = \frac{f''}{f} = 0.
\end{equation}
Equation~\eqref{eq:vk_z_eom} then reduces directly to the free wave equation for the vector potential on all scales\cite{Subramanian:2015lua,Kushwaha_2024},
\begin{equation}
A_k'' + k^2 A_k = 0.
\end{equation}
For super-Hubble modes ($k\eta \ll 1$), the general solution is\cite{Subramanian:2015lua,Kushwaha_2024}
\begin{equation}
A_k(\eta) = \mathcal{C}_1(k) + \mathcal{C}_2(k)\eta,
\label{eq:Ak_rd_solution}
\end{equation}
where integration constants $\mathcal{C}_1(k)$ and $\mathcal{C}_2(k)$ are determined by matching conditions across the non-singular bouncing cosmology\cite{PhysRevD.94.083524}. On super-horizon scales ($k\eta \ll 1$), the linear term $\mathcal{C}_2(k)\eta$ remains subdominant, and the comoving vector potential $A_k(\eta) \approx \mathcal{C}_1(k)$ freezes to a scale-dependent constant\cite{Subramanian:2015lua,Kushwaha_2024}.

While the physical magnetic energy density spectrum $\mathcal{P}_B(k,\eta)$ dilutes due to cosmic expansion as $a^{-4}(\eta)$, the comoving magnetic-field spectrum, defined by factoring out scale-factor expansion\cite{Subramanian:2015lua,Kushwaha_2024},
\begin{equation}
\mathcal{P}_{B,\mathrm{comoving}}(k,\eta) \equiv a^4(\eta) \mathcal{P}_B(k,\eta) = \frac{k^5 |A_k(\eta)|^2}{2\pi^2},
\label{eq:PB_comoving_def}
\end{equation}
is conserved on super-Hubble scales during radiation domination\cite{Subramanian:2015lua,Kushwaha_2024}:
\begin{equation}
\frac{d}{d\eta} \mathcal{P}_{B,\mathrm{comoving}}(k,\eta) \Big|_{k\eta \ll 1} = 0.
\end{equation}

The continuous matching of this amplitude across the bounce relies on two key physical assumptions \cite{PhysRevD.94.083524}:
\begin{enumerate}
    \item \textbf{Super-Horizon Scale Condition ($k\Delta\eta_{\mathrm{bounce}} \ll 1$):} The duration of the non-singular bouncing cosmology $\Delta\eta_{\mathrm{bounce}}$ is sufficiently brief compared to the mode wavelength such that relevant Fourier modes remain far outside the Hubble horizon throughout the entire transition\cite{Cai_2012,PhysRevD.94.083524}.
    \item \textbf{Rapid Coupling Decay:} The background scalar coupling $f(\phi)$ decays smoothly from $f_{\mathrm{end}} \sim \mathcal{O}(1)$ at the end of contraction to $f \to 1$ over a conformal time interval $\Delta\eta_{\mathrm{trans}} \lesssim \Delta\eta_{\mathrm{bounce}}$, preventing parametric excitation or amplification of sub-horizon vacuum modes\cite{Martin:2007ue,PhysRevD.94.083524}.
\end{enumerate}

Under these conditions, matching $A_k$ and its canonical momentum $\Pi_k \equiv f^2 A_k'$ across the non-singular bouncing cosmology confirms that the super-horizon amplitude $A_k(\eta_{\mathrm{end}})$ maps directly onto the post-bounce frozen mode $A_k^{\mathrm{RD}}$ without mode mixing or spurious particle production\cite{Membiela:2013cea,PhysRevD.94.083524,Kushwaha_2024}:
\begin{equation}
|A_k^{\mathrm{RD}}| \approx |A_k(\eta_{\mathrm{end}})|.
\end{equation}

Consequently, the comoving magnetic spectrum established at the end of contraction\cite{PhysRevD.94.083524},
\begin{equation}
\mathcal{P}_{B,\mathrm{comoving}}(k) = C_B k^{n_B}, \qquad C_B = \frac{1}{2\pi^2 f_{\mathrm{end}}^2} |a_{\mathrm{end}} H_{\mathrm{end}}|^{-n_B} H_{\mathrm{end}}^4,
\label{eq:PB_end_recap}
\end{equation}
is carried unchanged into the radiation-dominated era\cite{PhysRevD.94.083524,Kushwaha_2024}. This comoving spectrum can be applied directly to evaluate passive magnetic-induced curvature perturbations without requiring an auxiliary transfer function\cite{Giovannini:2000ch,Kushwaha_2024}.
%%%%%%%%%%%%%%%%%%%%%%%%%%%%%%%%%%%%%%%%%%%%%%%%%%%%%%%%%%%%
\section{ Curvature Power Spectrum: Numerical Evaluation and Analytical Scaling}
\label{sec:numerical_curvature}

\subsection{Conformal Time and Thermal Background Relations}
\label{subsec:conformal_time_thermal}

To track mode entry and evaluate super-horizon growth during the radiation domination, conformal time $\eta(T)$ is computed as a function of temperature $T$ by integrating over relativistic energy and entropy degrees of freedom ($g_*$ and $g_{*s}$) \cite{Saikawa_2018}:
\begin{equation}
\eta(T) = \frac{1}{a_{\mathrm{eq}} T_{\mathrm{eq}}} \sqrt{\frac{90}{8\pi^3 G}} \int_T^\infty \frac{dT'}{T'^2} \left( \frac{g_{*s}(T')}{g_{*s}(T_{\mathrm{eq}})} \right)^{1/3} \frac{1}{\sqrt{g_*(T')}} \left[ 1 + \frac{T'}{3g_{*s}} \frac{dg_{*s}}{dT'} \right],
\label{eq:conformal_time_integral}
\end{equation}
where $G = (8\pi M_{\mathrm{pl}}^2)^{-1}$ is Newton's gravitational constant. The scale factor at matter--radiation equality $a_{\mathrm{eq}}$ is defined as
\begin{equation}
a_{\mathrm{eq}} = \frac{4.15 \times 10^{-5}}{\Omega_m h^2},
\end{equation}
with baseline cosmological parameters $\Omega_m = 0.3106$ and $h = 0.677$\cite{2020A&A...641A...6P}.

\subsection{Scale Factor Normalization and $N$-Dependence of the Curvature Spectrum}
\label{subsec:scale_factor_normalization}

Entropy conservation ($g_{*s}(T) a^3 T^3 = \mathrm{constant}$) fixes the scale factor at the end of contraction $a_{\mathrm{end}}$ relative to the present CMB temperature $T_0 = 2.35 \times 10^{-13}~\mathrm{GeV}$\cite{Kolb:1990vq}:
\begin{equation}
a_{\mathrm{end}} = \frac{T_0}{T_{\mathrm{end}}} \left( \frac{g_{*s,0}}{g_{*s,\mathrm{end}}} \right)^{1/3}.
\label{eq:a_end_entropy}
\end{equation}

To consistently establish the scaling of the curvature perturbation amplitude with the contracting e-fold duration $N$, we analyze the full expression for the magnetic power spectrum prefactor $C_B$\cite{PhysRevD.94.083524}:
\begin{equation}
C_B = \frac{1}{2\pi^2 f_{\mathrm{end}}^2} |a_{\mathrm{end}} H_{\mathrm{end}}|^{-n_B} H_{\mathrm{end}}^4 = \frac{1}{2\pi^2 f_{\mathrm{end}}^2} a_{\mathrm{end}}^{-n_B} H_{\mathrm{end}}^{4 - n_B}.
\label{eq:CB_full_expression}
\end{equation}

In general, the quantities $(T_{\mathrm{end}}, H_{\mathrm{end}}, a_{\mathrm{end}})$ could be treated either as a dynamically linked family (in which a specific matching mechanism enforces $T_{\mathrm{end}} \propto H_{\mathrm{end}}^{1/2}$ and $a_{\mathrm{end}} \propto H_{\mathrm{end}}^{-1/2}$) or as a phenomenological family of contracting backgrounds sharing a fixed post-bounce end temperature. Throughout this work, we assume the latter model choice: we treat parameter scans in $N$ as a phenomenological family of background trajectories with a fixed post-bounce boundary temperature $T_{\mathrm{end}}$ (and consequently fixed $a_{\mathrm{end}}$ and $f_{\mathrm{end}} \equiv f(\phi(a_{\mathrm{end}}))$)\cite{PhysRevD.94.083524,Kushwaha_2024}. This choice cleanly decouples the background-contraction dynamics ($N, H_{\mathrm{end}}$) from model-dependent assumptions about the specific microphysics of the reheating transition.

Under this fixed end-state prescription ($a_{\mathrm{end}} = \mathrm{const}$, $f_{\mathrm{end}} = \mathrm{const}$), the $N$-dependence enters exclusively through the backreaction-constrained Hubble parameter $H_{\mathrm{end}}$\cite{PhysRevD.94.083524}:
\begin{equation}
H_{\mathrm{end}}(N) = \frac{\sqrt{2}\pi q}{5(1-q)} M_{\mathrm{pl}} e^{-N} \propto e^{-N}.
\label{eq:Hend_N_scaling}
\end{equation}

Substituting Eq.~\eqref{eq:Hend_N_scaling} into Eq.~\eqref{eq:CB_full_expression} yields the magnetic spectrum amplitude scaling\cite{PhysRevD.94.083524}:
\begin{equation}
C_B(N) \propto H_{\mathrm{end}}^{4 - n_B} \propto e^{-(4 - n_B)N}.
\end{equation}

Because passive curvature perturbations are quadratic in the magnetic field ($\zeta_B \propto \Pi_B \propto B^2$)\cite{Shaw:2009nf,Kushwaha_2024}, the curvature power spectrum prefactor $C_C$ scales with the square of the magnetic amplitude ($C_C \propto C_B^2$):
\begin{equation}
C_C(N) \propto C_B^2 \propto H_{\mathrm{end}}^{2(4 - n_B)} \propto e^{-2(4 - n_B)N}.
\label{eq:curvature_prefactor_scaling}
\end{equation}

For our benchmark magnetic spectral index $n_B \approx -0.212$, the exponent evaluates to $2(4 - n_B) = 2(4 - (-0.212)) = 8.424$. Consequently, under the fixed-$T_{\mathrm{end}}$ phenomenological family prescription, the curvature power spectrum amplitude exhibits an exponential scaling with $N$\cite{PhysRevD.94.083524,Kushwaha_2024}:
\begin{equation}
\mathcal{P}_\zeta(k) \propto e^{-8.42 N}.
\label{eq:Pzeta_N_scaling}
\end{equation}

For completeness, if one were instead to assume an instantaneous reheating scenario where $T_{\mathrm{end}} \propto H_{\mathrm{end}}^{1/2}$ (such that $a_{\mathrm{end}} \propto H_{\mathrm{end}}^{-1/2}$), $a_{\mathrm{end}}^{-n_B}$ would contribute an additional factor of $H_{\mathrm{end}}^{n_B/2}$, modifying the net exponent to $2(4 - n_B/2) \approx 8.212$. We adhere strictly to the fixed-$T_{\mathrm{end}}$ prescription ($e^{-8.42 N}$) across all benchmark scans in this work.

\subsection{Curvature Power Spectrum Sourced by Magnetic Anisotropic Stress}
\label{subsec:curvature_power_spectrum}

The anisotropic stress tensor of the magnetic field $\Pi_B(\mathbf{k},\eta)$ sources magnetic-field-induced curvature perturbations $\zeta_B(\mathbf{k},\eta)$ on super-Hubble scales ($k\eta \ll 1$)\cite{Shaw:2009nf,Bonvin:2010ze,Kushwaha_2024}:
\begin{equation}
\zeta_B(\mathbf{k},\eta) = -\frac{1}{3} \xi(\eta) R_\gamma(\eta) \Pi_B(\mathbf{k},\eta),
\label{eq:zeta_B_definition}
\end{equation}
where $R_\gamma(\eta) \equiv \rho_\gamma / \rho_{\mathrm{tot}}$ is the photon energy fraction. The auxiliary growth function $\xi(\eta)$ is parameterized by the scale factor ratio $a(\eta) / a_{\mathrm{end}}$\cite{Shaw:2009nf,Bonvin:2010ze,Kushwaha_2024}:
\begin{equation}
\xi(\eta) = \ln\left( \frac{a(\eta)}{a_{\mathrm{end}}} \right) + \frac{1}{2} \left( \frac{a_{\mathrm{end}}}{a(\eta)} \right) - \frac{1}{2}.
\label{eq:xi_scale_factor}
\end{equation}

Computing the two-point expectation value $\langle \zeta_B(\mathbf{k},\eta) \zeta_B^*(\mathbf{p},\eta) \rangle = (2\pi)^3 \delta^{(3)}(\mathbf{k}-\mathbf{p}) \frac{2\pi^2}{k^3} \mathcal{P}_{\zeta_B}(k,\eta)$ yields the dimensionless power spectrum\cite{PhysRevD.86.023512,Kushwaha_2024}:
\begin{equation}
\mathcal{P}_{\zeta_B}(k,\eta,T) = C_C(\eta,T) \frac{k^3}{2\pi^2} \int dk' \, k'^2 P_B(k') \int_{-1}^1 dg \, P_B(|\mathbf{k}-\mathbf{k}'|) \frac{\mathcal{N}(k,k',g)}{D(k,k',g)},
\label{eq:Pzeta_exact_integral}
\end{equation}
where $g \equiv \cos\theta = (\mathbf{k} \cdot \mathbf{k}') / (kk')$ is the cosine of the angle between interacting wavevectors. The thermal prefactor $C_C(\eta,T)$ is\cite{Kushwaha_2024}
\begin{equation}
C_C(\eta,T) = \frac{\xi^2(\eta) R_\gamma^2}{(2\pi)^3} \left( \frac{1}{16\pi \rho_{\gamma,0}^2} \right) \left( \frac{g_{*s}(T)}{g_{*s}(T_0)} \right)^{8/3} C_B^2,
\label{eq:CC_prefactor_def}
\end{equation}
with $C_B = \frac{1}{2\pi^2 f_{\mathrm{end}}^2} |a_{\mathrm{end}} H_{\mathrm{end}}|^{-n_B} H_{\mathrm{end}}^4$.

\subsection{Numerical Convolution and Kernel Evaluation}
\label{subsec:numerical_convolution}

The angular integral kernel $I_1(k,k')$ encapsulates the tensorial alignment of interacting modes\cite{PhysRevD.86.023512,Kushwaha_2024}:
\begin{equation}
I_1(k,k') = \int_{-1}^1 dg \, \mathcal{N}(k,k',g) \left[ D(k,k',g) \right]^{\frac{n_B}{2} - 1},
\label{eq:I1_kernel_def}
\end{equation}
where $D(k,k',g) \equiv k^2 + k'^2 - 2gkk'$ and the tensorial numerator $\mathcal{N}(k,k',g)$ is
\begin{equation}
\mathcal{N}(k,k',g) = \frac{1}{4} \left[ k^2(1 + g^2) + kk'(2g - 6g^3) + k'^2(5 + 9g^4 - 12g^2) \right].
\label{eq:N_kernel_def}
\end{equation}

When $k \approx k'$, $D(k,k',g)$ vanishes in the collinear limit $g \to 1$, inducing an integrable singularity for $n_B < 0$\cite{Kushwaha_2024}. To maintain double-precision accuracy without numerical instability, the angular domain is split at $g_0 = 0.9999$:
\begin{equation}
I_1(k,k') = \int_{-1}^{0.9999} dg \, \mathcal{N} D^{\frac{n_B}{2}-1} + \int_{0.9999}^{1 - \varepsilon_m} dg \, \mathcal{N} D^{\frac{n_B}{2}-1},
\label{eq:kernel_split}
\end{equation}
where $\varepsilon_m \approx 2.22 \times 10^{-16}$ is the machine epsilon. Convergence tests conducted across $g_0 \in [0.99, 0.99999]$ with adaptive quadrature tolerances $\varepsilon_{\mathrm{rel}} = 10^{-8}$ and $\varepsilon_{\mathrm{abs}} = 10^{-10}$ confirm numerical stability within a relative variation of $\Delta I_1 / I_1 < 10^{-6}$.

To avoid recalculating $I_1(k,k')$ inside the 2D momentum convolution across large parameter sweeps, $I_1(k,k')$ is tabulated on a 2D logarithmic grid\cite{Kushwaha_2024},
\begin{equation}
k_i \in [10^{-39}, 10^{-10}]~\mathrm{GeV}, \qquad k'_j \in [10^{-42}, 10^{-9}]~\mathrm{GeV},
\label{eq:lookup_grid}
\end{equation}
and interpolated using a 2D smooth bicubic spline $\mathcal{S}_{2\mathrm{D}}(x,x')$ on logarithmic coordinates $x \equiv \ln k$ and $x' \equiv \ln k'$. The accelerated kernel evaluation is $I_1^{\mathrm{fast}}(k,k') = \exp\left[ \mathcal{S}_{2\mathrm{D}}(\ln k, \ln k') \right]$.

Transforming the momentum variable to $x' \equiv \ln k'$, the curvature spectrum at horizon-crossing conformal time $\eta_*(k) \equiv \tilde{\eta}_{\mathrm{end}} + \tilde{q}/k$ is evaluated using adaptive quadrature\cite{Kushwaha_2024}:
\begin{equation}
\mathcal{P}_\zeta^{\mathrm{HC}}(k) = C_C(\eta_*(k), T) \frac{k^3}{2\pi^2} \int_{\ln k_{\min}}^{\ln k_{\max}} dx' \, e^{(3 + n_B)x'} I_1^{\mathrm{fast}}\left(k, e^{x'}\right).
\label{eq:Pzeta_HC_integral}
\end{equation}

\begin{figure}[htbp]
\centering
\includegraphics[width=0.75\textwidth]{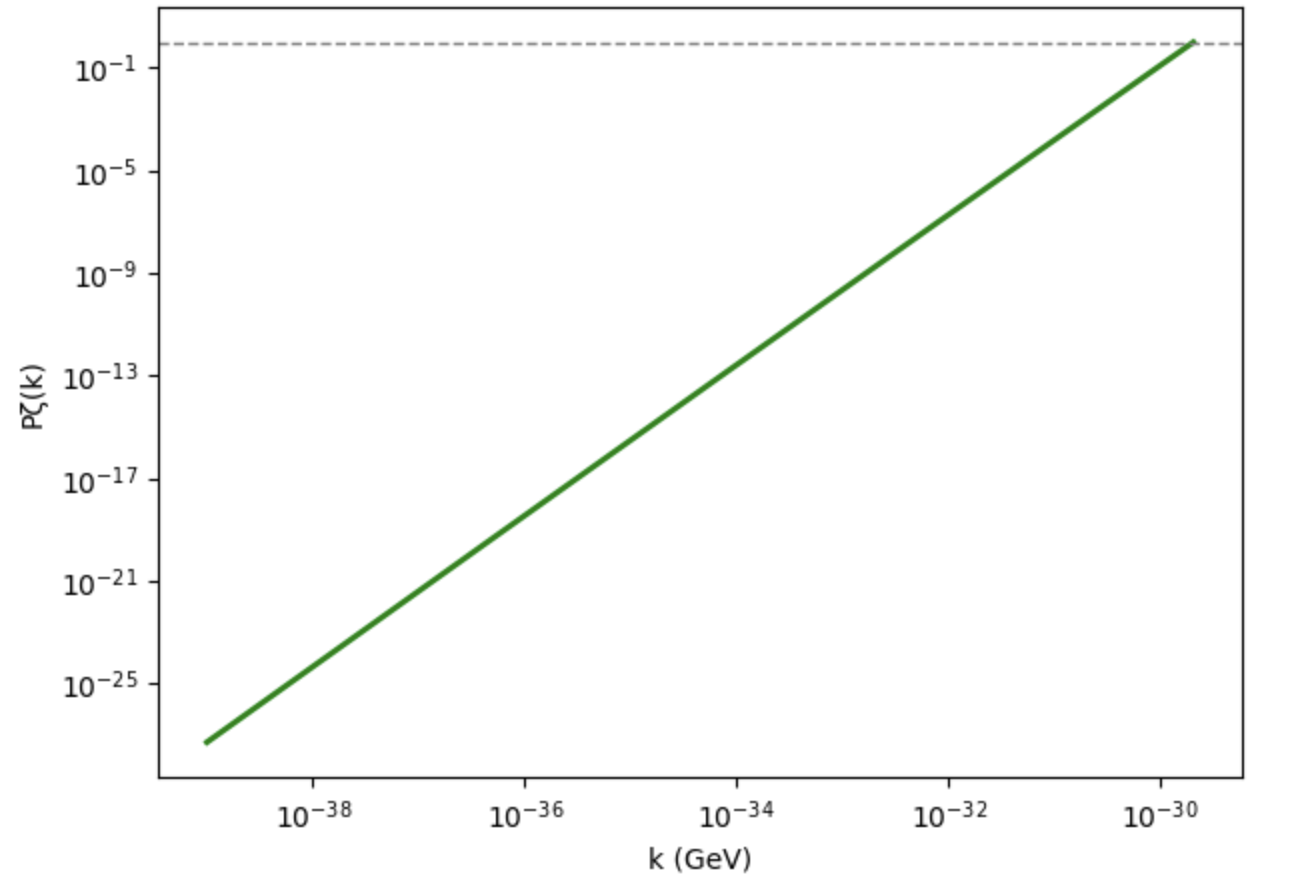}
\caption{The horizon-crossing primordial curvature power spectrum $\mathcal{P}_\zeta^{\mathrm{HC}}(k)$ evaluated numerically using fixed physical wavenumber cutoffs $k_{\min} = 10^{-39}~\mathrm{GeV}$ and $k_{\max} = 10^{-30}~\mathrm{GeV}$ for benchmark parameters $q = 0.237$, $N = 33.74$, and $H_{\mathrm{end}} = 1493.901~\mathrm{GeV}$. The spectrum exhibits the characteristic blue-tilted slope predicted by the UV-dominated convolution, reaching $\mathcal{P}_\zeta^{\mathrm{HC}} \sim \mathcal{O}(10^{-1})$ at the numerical wavenumber cutoff $k_{\mathrm{num,max}} = 10^{-30}~\mathrm{GeV} \ll k_{\mathrm{SH,max}} \approx 2.2 \times 10^{-28}~\mathrm{GeV}$, preserving super-Hubble validity.}
\label{fig:power_spectrum_curve}
\end{figure}

Super-Hubble validity requires $|k\eta_{\mathrm{end}}| \ll 1$, giving an upper bound $k_{\mathrm{SH,max}} \approx a_{\mathrm{end}} H_{\mathrm{end}} \approx 2.2 \times 10^{-28}~\mathrm{GeV}$ for $H_{\mathrm{end}} = 1493.901~\mathrm{GeV}$\cite{PhysRevD.94.083524,Kushwaha_2024}. To ensure super-Hubble conditions hold throughout the integration, we set a conservative numerical wavenumber cutoff $k_{\mathrm{num,max}} = 10^{-30}~\mathrm{GeV} \ll k_{\mathrm{SH,max}}$.

\subsection{Analytical Derivation of the Spectral Tilt and PBH Formation Implication}
\label{subsec:analytical_spectral_tilt}

We assume the logarithmic spectral tilt of the curvature power spectrum as \cite{Liddle:1992wi}
\begin{equation}
n_\zeta(k) \equiv \frac{d\ln \mathcal{P}_\zeta^{\mathrm{HC}}}{d\ln k}.
\label{eq:tilt_definition}
\end{equation}
To prevent notation collisions, we reserve $n_\zeta$ strictly for the spectral index and $\eta$ for conformal time ($\eta_*$).

\subsubsection{Physical Origin of UV Dominance}
The physical origin of the spectral shape lies in the quadratic nature of the vector-field energy-momentum tensor\cite{PhysRevD.86.023512,Kushwaha_2024}. Magnetic anisotropic stress is quadratic in the vector field ($\Pi_{ij} \propto B_i B_j$). Consequently, a curvature perturbation mode at wavenumber $k$ is sourced through a momentum convolution over mode pairs ($\mathbf{k}', \mathbf{k}-\mathbf{k}'$). For spectral indices $n_B > -1.5$, the magnetic power spectrum grows with wavenumber, concentrating energy at small spatial scales. As a result, high-frequency mode interactions dominate the momentum integral. This concentration renders the convolution inherently UV-dominated near the physical wavenumber cutoff $k_{\max}$ \cite{PhysRevD.86.023512}.

\subsubsection{Asymptotic Spectral Tilt and Fixed UV Cutoff}
To quantify this physical mechanism analytically, we assume the scale-free benchmark ($k_{\min} \to 0, k_{\max} \to \infty$). We substitute $x \equiv k'/k$ into Eq.~\eqref{eq:Pzeta_HC_integral}. This transformation makes the integral scale-independent and yields \cite{PhysRevD.86.023512}
\begin{equation}
\mathcal{P}_\zeta^{\mathrm{HC}}(k) \propto C_C(\eta_*(k)) k^{6 + 2n_B} \implies n_\zeta^{\mathrm{scale-free}} = 6 + 2n_B + \frac{d\ln C_C}{d\ln k}.
\label{eq:Pzeta_scale_free}
\end{equation}
For $n_B = -0.212$, the scale-free exponent evaluates to $6 + 2n_B \approx 5.576$.

Realistic physical setups feature bounded integration limits ($k_{\min}, k_{\max}$). In this case, the upper wavenumber cutoff $k_{\max}$ restricts the integration domain of the convolution. The dimensionless convolution integral $J(k)$ depends explicitly on $k$ through its upper integration boundary\cite{Kushwaha_2024}:
\begin{equation}
J(k) = \int_{k_{\min}/k}^{k_{\max}/k} dx \, x^{2+n_B} I(x,n_B).
\label{eq:J_k_def}
\end{equation}
We assume the asymptotic regime $k_{\min} \ll k \ll k_{\max}$. Here, the integrand is dominated by its upper limit ($x \gg 1$), where $I(x) \approx 1.4 x^{n_B}$. Integrating over this UV tail yields $J(k) \propto (k_{\max}/k)^{3 + 2n_B}$. This boundary scaling suppresses the power spectrum at small wavenumbers relative to $k_{\max}$, yielding\cite{PhysRevD.86.023512,Kushwaha_2024}
\begin{equation}
\frac{d\ln J(k)}{d\ln k} \approx -\left(3 + 2n_B\right).
\label{eq:dlnJ_dlnk}
\end{equation}

Combining the scale-free convolution tilt with the cutoff boundary contribution isolates the net spectral tilt\cite{Kushwaha_2024}:
\begin{equation}
n_\zeta(k) = \underbrace{(6 + 2n_B)}_{\Delta n_\zeta^{\mathrm{Conv}}} + \underbrace{\left( -(3 + 2n_B) \right)}_{\Delta n_\zeta^{\mathrm{Cutoff}}} + \underbrace{\frac{d\ln C_C}{d\ln k}}_{\Delta n_\zeta^{\mathrm{HC}}} = 3 + \Delta n_\zeta^{\mathrm{HC}}.
\label{eq:tilt_cancellation_step}
\end{equation}

Equation~\eqref{eq:tilt_cancellation_step} demonstrates that the $2n_B$ magnetic index dependence cancels at leading order. This cancellation establishes a blue-tilted, asymptotic $k^3$ scaling for the curvature power spectrum inside the band $k \ll k_{\max}$\cite{PhysRevD.86.023512,Kushwaha_2024}. 

Next, we evaluate the horizon-crossing correction\cite{Kushwaha_2024}:
\begin{equation}
\Delta n_\zeta^{\mathrm{HC}} = -\frac{2\tilde{q}}{k \eta_* \xi_*} \left( 1 - \frac{\eta_B}{2\eta_*} \right) \approx -\frac{2}{\xi_*} \approx -0.20.
\label{eq:horizon_crossing_correction}
\end{equation}
With this correction, the localized spectral tilt evaluates to $n_\zeta \approx 2.80$. This analytical slope is fully confirmed by the numerical quadrature shown in Figure~\ref{fig:power_spectrum_curve}.

\subsubsection{Density Variance Integration and Implication for PBHs}
The blue-tilted curvature power spectrum $\mathcal{P}_\zeta^{\mathrm{HC}}(k) \propto k^{2.80}$ feeds directly into the smoothed density variance $\sigma_B^2(R)$\cite{Young:2014ana,Kushwaha_2024}. In Fourier space, this calculation integrates $\mathcal{P}_\zeta^{\mathrm{HC}}(k)$ against a Gaussian window function $W(k'R)$\cite{Green:2004wb,Young:2019bvu}. The window function exponentially suppresses contributions from modes smaller than the smoothing scale $R = k^{-1}$\cite{Harada:2013epa}.

Because the underlying curvature spectrum rises sharply toward small scales, the density variance accumulates power up to the cutoff scale $R \approx k_{\max}^{-1}$\cite{PhysRevD.86.023512,Kushwaha_2024}. The variance reaches its maximum amplitude $\sigma_B^2(R)$ near this smoothing scale.

This localized peak in the density variance increases the probability of large fluctuations in density contrast \cite{Young:2014ana, Sasaki_2018}. Overdensities on scales near $R \approx k_{\max}^{-1}$ can exceed the critical collapse threshold $\delta_c$ upon horizon re-entry during radiation domination\cite{Harada:2013epa}. Consequently, this threshold traversal triggers efficient PBH formation \cite{Green_2021, Kushwaha_2024}. It fixes the primary horizon mass scale and determines the peak amplitude of the black hole mass abundance\cite{Sasaki_2018,Carr_2021}.

%%%%%%%%%%%%%%%%%%%%%%%%%%%%%%%%%%%%%%%%%%%%%%%%%%%%%%%%%%%%
\section{Variance of Density Perturbations}
\label{sec:density_variance}

To evaluate the initial abundance of PBHs formed during radiation domination\cite{Sasaki_2018}, we calculate the smoothed mass variance of magnetic-field-induced density overdensities on a spatial scale $R$\cite{Young:2014ana,Kushwaha_2024}. The spatial smoothing scale $R$ is related to the characteristic comoving Fourier wavenumber by $R = k^{-1}$. In Fourier space, this smoothing is implemented using a standard Gaussian filter\cite{Green:2004wb,Young:2019bvu},
\begin{equation}
W(k'R) = e^{-k'^2 R^2 / 2}.
\label{eq:gaussian_window}
\end{equation}

During radiation domination, linear density contrast fluctuations $\delta(\mathbf{k},\eta)$ on sub-horizon scales are linked to comoving curvature perturbations $\zeta(\mathbf{k})$ via the Poisson relation $\delta(\mathbf{k},\eta) = \frac{4}{9}(k / aH)^2 \zeta(\mathbf{k})$\cite{Harada:2013epa,Young:2014ana}. At the moment of horizon re-entry ($k = aH$), this expression reduces to $\delta \approx \frac{4}{9}\zeta$. The smoothed density variance $\sigma_B^2(R)$ is obtained by integrating over the dummy momentum variable $k'$ of the dimensionless curvature power spectrum evaluated at horizon crossing, $\mathcal{P}_\zeta^{\mathrm{HC}}(k')$\cite{Young:2014ana,Kushwaha_2024}:
\begin{equation}
\sigma_B^2(R) = \frac{16}{81} \int_{k_{\min}}^{k_{\max}} \frac{dk'}{k'} (k'R)^4 W^2(k'R) \mathcal{P}_\zeta^{\mathrm{HC}}(k').
\label{eq:variance_density}
\end{equation}

We evaluate the power spectrum at horizon crossing in Eq.~\eqref{eq:variance_density} because passive curvature perturbations sourced by magnetic anisotropic stresses grow cumulatively on super-Hubble scales until horizon entry ($k'\eta_* \sim 1$)\cite{Shaw:2009nf,Bonvin:2010ze,Kushwaha_2024}. Evaluating the spectrum at horizon crossing captures the maximum linear amplitude attained by each mode before sub-horizon pressure oscillations and radiative dissipation. This choice provides the amplitude used in the adopted calculation required for collapse threshold calculations \cite{Harada:2013epa, Kushwaha_2024}.

\begin{figure}[htbp]
\centering
\includegraphics[width=0.75\textwidth]{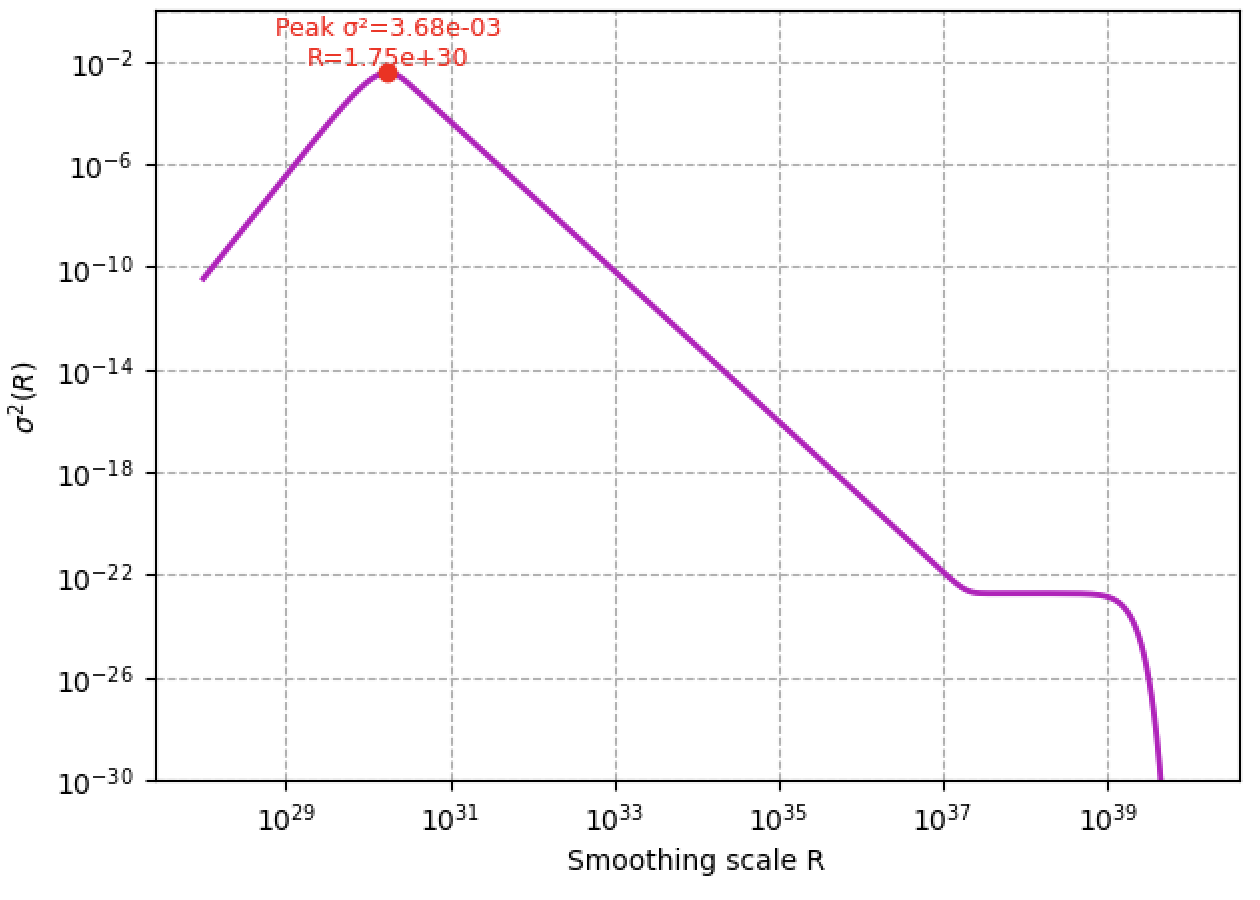}
\caption{Smoothed density variance $\sigma_B^2(R)$ as a function of spatial smoothing scale $R = k^{-1}$, calculated using a Gaussian window function for baseline parameters $q = 0.237$, $N = 33.74$, and $H_{\mathrm{end}} = 1493.901~\mathrm{GeV}$\cite{Young:2014ana,Kushwaha_2024}. The input power spectrum is integrated over $k' \in [10^{-39}, 10^{-30}]~\mathrm{GeV}$. The variance reaches a peak amplitude of $\sigma_B^2 \approx 3.68 \times 10^{-3}$ at $R \approx 1.76 \times 10^{30}~\mathrm{GeV}^{-1}$, determining the primary scale for PBH production.}
\label{fig:sigma2_vs_R}
\end{figure}

Figure~\ref{fig:sigma2_vs_R} illustrates the resulting smoothed density variance $\sigma_B^2(R)$ as a function of the spatial smoothing scale $R = k^{-1}$ for baseline parameters $q = 0.237$, $N = 33.74$, and $H_{\mathrm{end}} = 1493.901~\mathrm{GeV}$. Integrating the input power spectrum over $k' \in [10^{-39}, 10^{-30}]~\mathrm{GeV}$ yields a peak variance of $\sigma_B^2 \approx 3.68 \times 10^{-3}$ at $R \approx 1.76 \times 10^{30}~\mathrm{GeV}^{-1}$. This peak scale determines the primary horizon scale for PBH production.

\begin{figure}[htbp]
\centering
\includegraphics[width=0.75\textwidth]{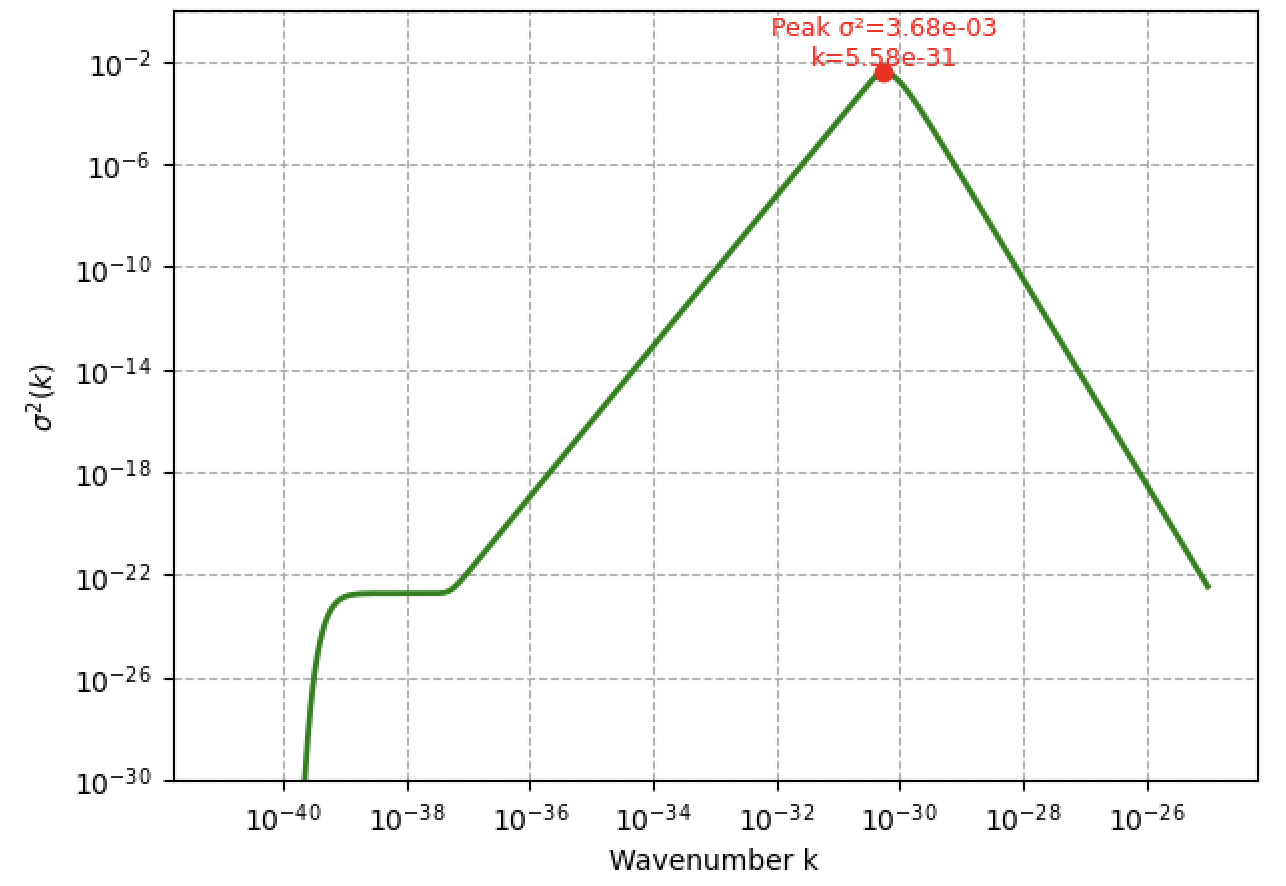}
\caption{Density variance $\sigma_B^2(k)$ plotted as a function of the evaluation wavenumber scale $k \equiv R^{-1}$ for benchmark parameters $q = 0.237$, $N = 33.74$, and $H_{\mathrm{end}} = 1493.901~\mathrm{GeV}$\cite{Young:2014ana,Kushwaha_2024}. We explicitly distinguish the input spectrum integration domain $k' \in [10^{-39}, 10^{-30}]~\mathrm{GeV}$ from the plotted range of evaluation wavenumbers $k$, which extends up to $10^{-25}~\mathrm{GeV}$ to illustrate the rapid exponential damping enforced by the Gaussian filter $W(k'R)$ at scales smaller than the spectrum cutoff $k_{\max} = 10^{-30}~\mathrm{GeV}$. The spectrum reaches a maximum value of $\sigma_B^2 \approx 3.68 \times 10^{-3}$ at $k \approx 5.58 \times 10^{-31}~\mathrm{GeV}$.}
\label{fig:sigma2_vs_k}
\end{figure}

Alternatively, Figure~\ref{fig:sigma2_vs_k} presents the density variance $\sigma_B^2(k)$ as a function of the evaluation wavenumber scale $k \equiv R^{-1}$. Here, we explicitly distinguish the input spectrum integration domain $k' \in [10^{-39}, 10^{-30}]~\mathrm{GeV}$ from the evaluation wavenumber $k$. Plotting up to $k = 10^{-25}~\mathrm{GeV}$ demonstrates the rapid exponential damping enforced by the Gaussian filter $W(k'R)$ at scales smaller than the spectrum wavenumber cutoff $k_{\max} = 10^{-30}~\mathrm{GeV}$. The spectrum reaches a maximum value of $\sigma_B^2 \approx 3.68 \times 10^{-3}$ at $k \approx 5.58 \times 10^{-31}~\mathrm{GeV}$.

%%%%%%%%%%%%%%%%%%%%%%%%%%%%%%%%%%%%%%%%%%%%%%%%%%%%%%%%%%%%
\newpage
\section{Framework for PBH Formation }
\label{sec:pbh_framework}

In this section, we find the physical framework governing PBH formation during radiation domination from magnetic-field-induced curvature perturbations. We assume a thermal mapping that connects comoving wavenumbers to parent horizon-entry temperatures and horizon masses, incorporate scale-dependent threshold corrections, and define the initial collapse fraction.

\subsection{Thermal Mapping: Wavenumber, Horizon Re-entry, and Horizon Mass}
\label{subsec:thermal_mapping}

During radiation domination, the horizon mass $M_H(T)$ enclosed within a Hubble radius $d_H = H^{-1}(T)$ at background temperature $T$ is defined as\cite{Green:2004wb,Sasaki_2018}
\begin{equation}
M_H(T) = \frac{4\pi M_{\mathrm{pl}}^2}{H(T)},
\label{eq:MH_definition}
\end{equation}
where $M_{\mathrm{pl}} = 2.435 \times 10^{18}~\mathrm{GeV}$ is the reduced Planck mass and $H(T)$ accounts for relativistic energy degrees of freedom $g_*(T)$\cite{Saikawa_2018}:
\begin{equation}
H(T) = \sqrt{\frac{\pi^2 g_*(T)}{90}} \frac{T^2}{M_{\mathrm{pl}}}.
\label{eq:H_temperature}
\end{equation}

A comoving mode $k$ re-enters the Hubble horizon during radiation domination at background temperature $T_f$, defined strictly by the horizon-crossing condition\cite{Harada:2013epa,Sasaki_2018}
\begin{equation}
k = a(T_f) H(T_f).
\label{eq:horizon_crossing_cond}
\end{equation}
We emphasize that $T_f$ specifically represents the background temperature at the moment when the parent comoving spatial scale $R = k^{-1}$ re-enters the horizon. It is not automatically identical to the instantaneous temperature at black hole formation or during fluid collapse, which can involve finite collapse-time delays or critical-scaling dynamics during sub-horizon evolution.

Applying entropy conservation ($g_{*s}(T) a^3 T^3 = \mathrm{constant}$)\cite{Saikawa_2018}, the scale factor at horizon re-entry $a(T_f)$ relates to matter--radiation equality ($a_{\mathrm{eq}}, T_{\mathrm{eq}}$) via
\begin{equation}
a(T_f) = a_{\mathrm{eq}} \left( \frac{T_{\mathrm{eq}}}{T_f} \right) \left( \frac{g_{*s}(T_{\mathrm{eq}})}{g_{*s}(T_f)} \right)^{1/3},
\label{eq:scale_factor_entropy}
\end{equation}
where $g_{*s}(T)$ denotes effective entropy degrees of freedom\cite{Saikawa_2018}. Multiplying Eq.~\eqref{eq:scale_factor_entropy} by $H(T_f)$ yields the explicit horizon-entry relation mapping comoving wavenumber $k$ to the re-entry temperature $T_f$\cite{Green:2004wb,Carr_2021}:
\begin{equation}
k(T_f) = a_{\mathrm{eq}} T_{\mathrm{eq}} \left( \frac{g_{*s}(T_{\mathrm{eq}})}{g_{*s}(T_f)} \right)^{1/3} \frac{H(T_f)}{T_f}.
\label{eq:k_Tf_mapping}
\end{equation}

For temperatures $T_f \in [10^{-8}, 10^{12}]~\mathrm{GeV}$, Eq.~\eqref{eq:k_Tf_mapping} is inverted numerically to map each Fourier wavenumber $k$ to its parent horizon re-entry temperature $T_f$ and enclosed horizon mass $M_H(T_f)$.

\subsection{Scale-Dependent Collapse Threshold and Domain of Validity}
\label{subsec:collapse_threshold}

The magnetic growth parameter $\xi_{\mathrm{RD}}(k)$ is evaluated at horizon entry. It depends on the scale factor ratio $x_f(k) \equiv a_{\mathrm{end}} / a(T_f)$ between the end of ekpyrotic contraction ($T_{\mathrm{end}}$) and mode re-entry ($T_f$). We apply entropy conservation ($a \propto T^{-1} g_{*s}^{-1/3}$). Under this condition, $x_f(k)$ evaluates to
\begin{equation}
x_f(k) = \left( \frac{T_f(k)}{T_{\mathrm{end}}} \right) \left( \frac{g_{*s}(T_f(k))}{g_{*s}(T_{\mathrm{end}})} \right)^{1/3}.
\label{eq:xf_correct_ratio}
\end{equation}
This ratio yields the scale-dependent cumulative growth function:
\begin{equation}
\xi_{\mathrm{RD}}(k) = \ln\left( x_f(k) \right) + \frac{1}{2 x_f(k)} - \frac{1}{2}.
\label{eq:xi_RD_definition}
\end{equation}

The background magnetic anisotropic stress modifies the fluid pressure response. As a result, the critical overdensity threshold $\delta_B(k)$ required for gravitational collapse receives scale-dependent corrections. In terms of the radiation sound speed $c_s^2 = 1/3$, the collapse threshold is given by\cite{Kushwaha_2024}
\begin{equation}
\delta_B(k) = \frac{c_s^2}{1 - \dfrac{5.8}{\xi_{\mathrm{RD}}(k)}}.
\label{eq:delta_B_th}
\end{equation}

\subsubsection*{Physical Domain of Validity}

Equation~\eqref{eq:delta_B_th} possesses a mathematical singularity at $\xi_{\mathrm{RD}} = 5.8$. Furthermore, it turns negative for $\xi_{\mathrm{RD}} < 5.8$. Consequently, the physical domain of validity for this threshold prescription is strictly restricted to
\begin{equation}
\xi_{\mathrm{RD}}(k) > 5.8.
\label{eq:threshold_domain}
\end{equation}

We verify that $\delta_B(k)$ remains well-behaved and physically meaningful throughout our calculations. For this purpose, we evaluate $\xi_{\mathrm{RD}}(k)$ across the active parent-scale range $k \in [k_{\min}, k_{\max}]$:
\begin{itemize}
    \item We assume an end-of-contraction scale $T_{\mathrm{end}} \approx 3.3 \times 10^{10}~\mathrm{GeV}$ and parent horizon-entry temperatures $T_f \in [10^{-8}, 10^2]~\mathrm{GeV}$. Here, modes re-enter during radiation domination after the non-singular bouncing cosmology ($T_f \ll T_{\mathrm{end}}$).
    \item The resulting scale-factor ratio is extremely small, spanning $10^{-19} \lesssim x_f(k) \lesssim 10^{-8}$.
    \item For $x_f(k) \ll 1$, the linear inverse term in Eq.~\eqref{eq:xi_RD_definition} dominates. This places the growth parameter in the range $10^7 \lesssim \xi_{\mathrm{RD}}(k) \lesssim 10^{18}$.
\end{itemize}

Across the entire parent-scale spectrum, $\xi_{\mathrm{RD}}(k) \ge 10^7 \gg 5.8$. Thus, $\delta_B(k)$ operates strictly within its valid domain. The threshold remains well-defined, strictly positive, and bounded:
\begin{equation}
\frac{1}{3} < \delta_B(k) \le 0.333331.
\end{equation}
It relaxes rapidly toward the standard value in the radiation-dominated era, $\delta_c = c_s^2 = 1/3$ \cite{Harada:2013epa}, without encountering unphysical singularities.

\subsection{Initial Collapse Fraction}
\label{subsec:initial_collapse_fraction}

Assuming a Gaussian probability distribution for smoothed overdensities $\delta$ on a spatial smoothing scale $R = k^{-1}$, the fraction $\beta(R)$ of horizon volumes collapsing into PBH at formation is given by the Gaussian tail integral\cite{Green:2004wb,Young:2019bvu,Sasaki_2018}:
\begin{equation}
\beta(R) = \frac{\gamma_c}{\sqrt{2\pi} \sigma_B(R)} \int_{\delta_B(k)}^{\infty} \exp\left( -\frac{\delta^2}{2\sigma_B^2(R)} \right) d\delta = \frac{\gamma_c}{2} \, \mathrm{erfc}\left( \frac{\delta_B(k)}{\sqrt{2}\sigma_B(R)} \right),
\label{eq:beta_exact_erfc}
\end{equation}
where $\gamma_c \approx 0.2$ parameterizes the efficiency of gravitational collapse during radiation domination\cite{Harada:2013epa}, $\sigma_B(R) \equiv \sqrt{\sigma_B^2(R)}$ is the root-mean-square mass variance evaluated with a Gaussian window function at $R = k^{-1}$, and $\mathrm{erfc}(x)$ is the complementary error function.

In the rare-event regime where the threshold-to-variance ratio satisfies $\nu(R) \equiv \delta_B(k)/\sigma_B(R) \gg 1$, we apply the standard asymptotic expansion $\mathrm{erfc}(x) \sim \frac{e^{-x^2}}{\sqrt{\pi}x} \left(1 - \frac{1}{2x^2} + \dots \right)$ to evaluate Eq.~\eqref{eq:beta_exact_erfc}. Truncating at leading order yields the rare-event asymptotic approximation\cite{Green:2004wb,Sasaki_2018}:
\begin{equation}
\beta(R) \approx \frac{\gamma_c}{\sqrt{2\pi}} \left( \frac{\sigma_B(R)}{\delta_B(k)} \right) \exp\left( -\frac{\delta_B^2(k)}{2\sigma_B^2(R)} \right).
\label{eq:beta_gaussian_asymptotic}
\end{equation}

\section{PBH Mass Abundance: Monochromatic and Critical Collapse Approaches}
\label{sec:abundance_formalisms}

In this section, we find two complementary formalisms for evaluating the present-day PBH mass abundance: (i) the monochromatic approximation, and (ii) the continuous critical-collapse mass function. For clarity, we distinguish our derivations of abundance. We explicitly distinguish between three key physical metrics:
\begin{enumerate}
    \item \textbf{Initial Collapse Fraction $\beta(R)$:} The dimensionless energy fraction of the Universe collapsing into PBHs at the moment of horizon re-entry ($\eta = \eta_f$)\cite{Green:2004wb}.
    \item \textbf{Monochromatic Present-Day Abundance $f_{\mathrm{PBH}}^{\mathrm{mono}}(M)$:} The ratio of present-day PBH mass density to total dark matter density ($\Omega_{\mathrm{PBH}}/\Omega_{\mathrm{DM}}$) calculated under the assumption that all overdensities on scale $R$ collapse into black holes of a single fixed mass $M = \gamma_c M_H$\cite{PhysRevD.81.104019,Sasaki_2018,Carr_2021}.
    \item \textbf{Differential Critical-Collapse Abundance $\dfrac{d f_{\mathrm{PBH}}}{d \ln M}$:} The continuous spectral distribution of present-day PBH dark matter fraction per logarithmic mass interval, incorporating near-threshold critical scaling ($M = K M_H (\delta - \delta_c)^\gamma$)\cite{Byrnes_2018}.
\end{enumerate}

\subsection{Formalism I: Monochromatic Scale-to-Mass Mapping}
\label{subsec:monochromatic_formalism}

The monochromatic approximation assumes a direct, one-to-one correspondence between the parent comoving scale $R = k^{-1}$ re-entering the horizon and the final black hole mass $M$. In this limit, an overdense region collapsing at horizon entry produces a PBH with mass proportional to the enclosed horizon mass\cite{Harada:2013epa}:
\begin{equation}
M = \gamma_c M_H(T_f),
\label{eq:monochromatic_mass_relation}
\end{equation}
where $\gamma_c \approx 0.2$ is the monochromatic collapse efficiency factor\cite{Harada:2013epa}, $M_H(T_f)$ denotes the total enclosed horizon mass at background temperature $T_f$, and $M$ represents the resulting PBH mass.

The initial collapse fraction $\beta(R)$, defined as the ratio of PBH mass density to total background density at formation, is given by Press--Schechter integration over the Gaussian tail\cite{Green:2004wb,Sasaki_2018}:
\begin{equation}
\beta(R) \equiv \left. \frac{\rho_{\mathrm{PBH}}(R)}{\rho_{\mathrm{tot}}} \right|_{\mathrm{formation}} = \frac{\gamma_c}{\sqrt{2\pi}} \left( \frac{\sigma_B(R)}{\delta_B(k)} \right) \exp\left( -\frac{\delta_B^2(k)}{2\sigma_B^2(R)} \right).
\label{eq:beta_initial_def}
\end{equation}

Redshifting the radiation energy density relative to cold dark matter from formation ($T_f$) until matter--radiation equality converts the initial collapse fraction $\beta(M)$ into the present-day monochromatic dark matter energy fraction $f_{\mathrm{PBH}}^{\mathrm{mono}}(M) \equiv \Omega_{\mathrm{PBH}}(M) / \Omega_{\mathrm{DM}}$\cite{PhysRevD.81.104019}:
\begin{equation}
f_{\mathrm{PBH}}^{\mathrm{mono}}(M) = \frac{\beta(M)}{3.27 \times 10^{-8}} \left( \frac{106.75}{g_{*,f}} \right)^{1/4} \left( \frac{M}{M_\odot} \right)^{-1/2},
\label{eq:fpbh_monochromatic}
\end{equation}
where $M$ is expressed in units of solar mass $M_\odot \approx 1.989 \times 10^{30}~\mathrm{kg}$, and $g_{*,f}$ represents the relativistic degrees of freedom at formation\cite{Saikawa_2018}. The corresponding present-day monochromatic PBH density parameter is
\begin{equation}
\Omega_{\mathrm{PBH}}^{\mathrm{mono}}(M) = f_{\mathrm{PBH}}^{\mathrm{mono}}(M) \Omega_{\mathrm{DM},0},
\label{eq:Omega_PBH_definition}
\end{equation}
with baseline dark matter density $\Omega_{\mathrm{DM},0} = 0.265$.

\begin{figure}[htbp]
    \centering
    \begin{minipage}[b]{0.6\textwidth}
        \centering
        \includegraphics[width=\textwidth]{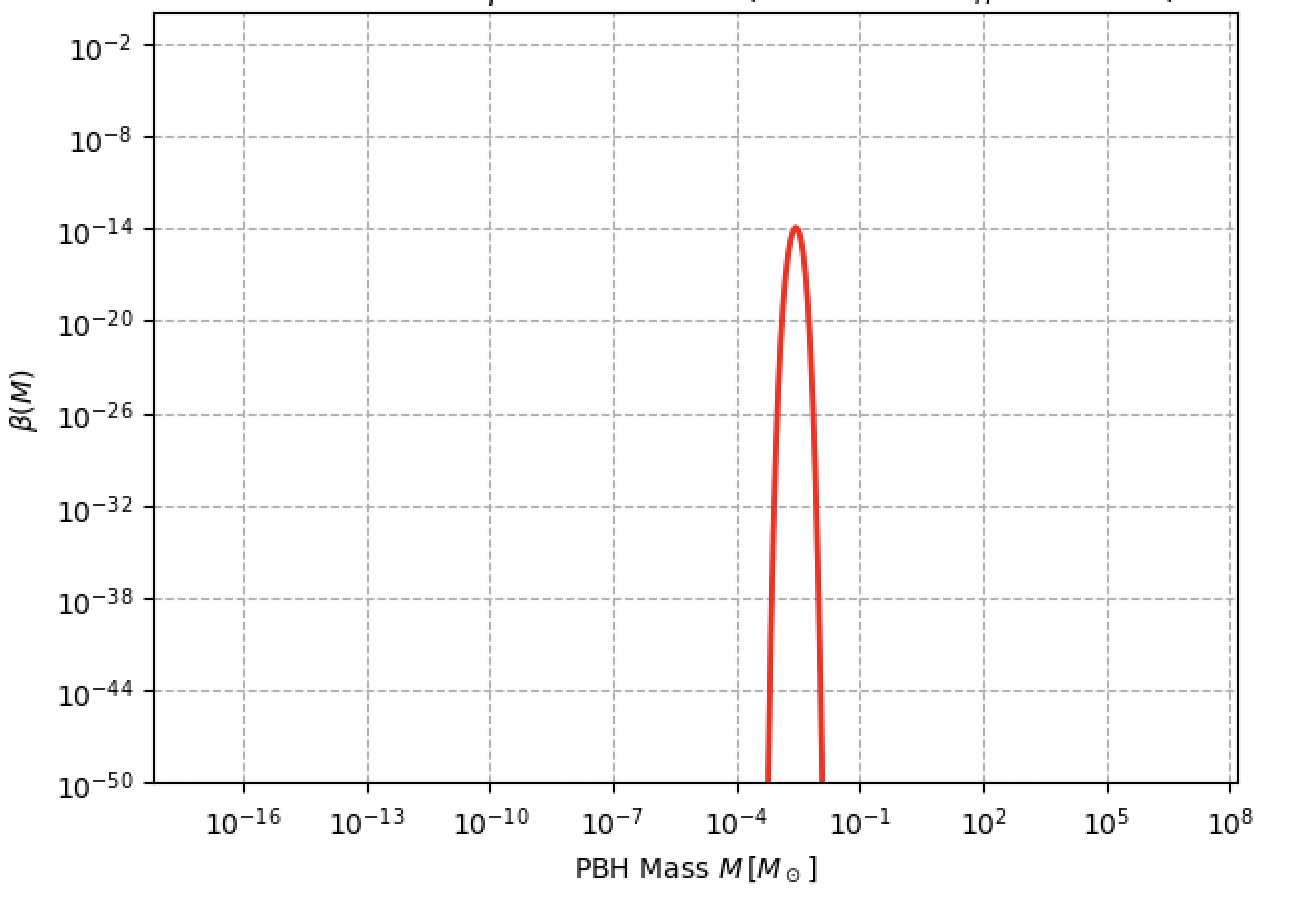}
        \subcaption{Initial collapse fraction $\beta(M)$.}
        \label{fig:beta_mass}
    \end{minipage}
    \hspace{0.2cm}
    \begin{minipage}[b]{0.6\textwidth}
        \centering
        \includegraphics[width=\textwidth]{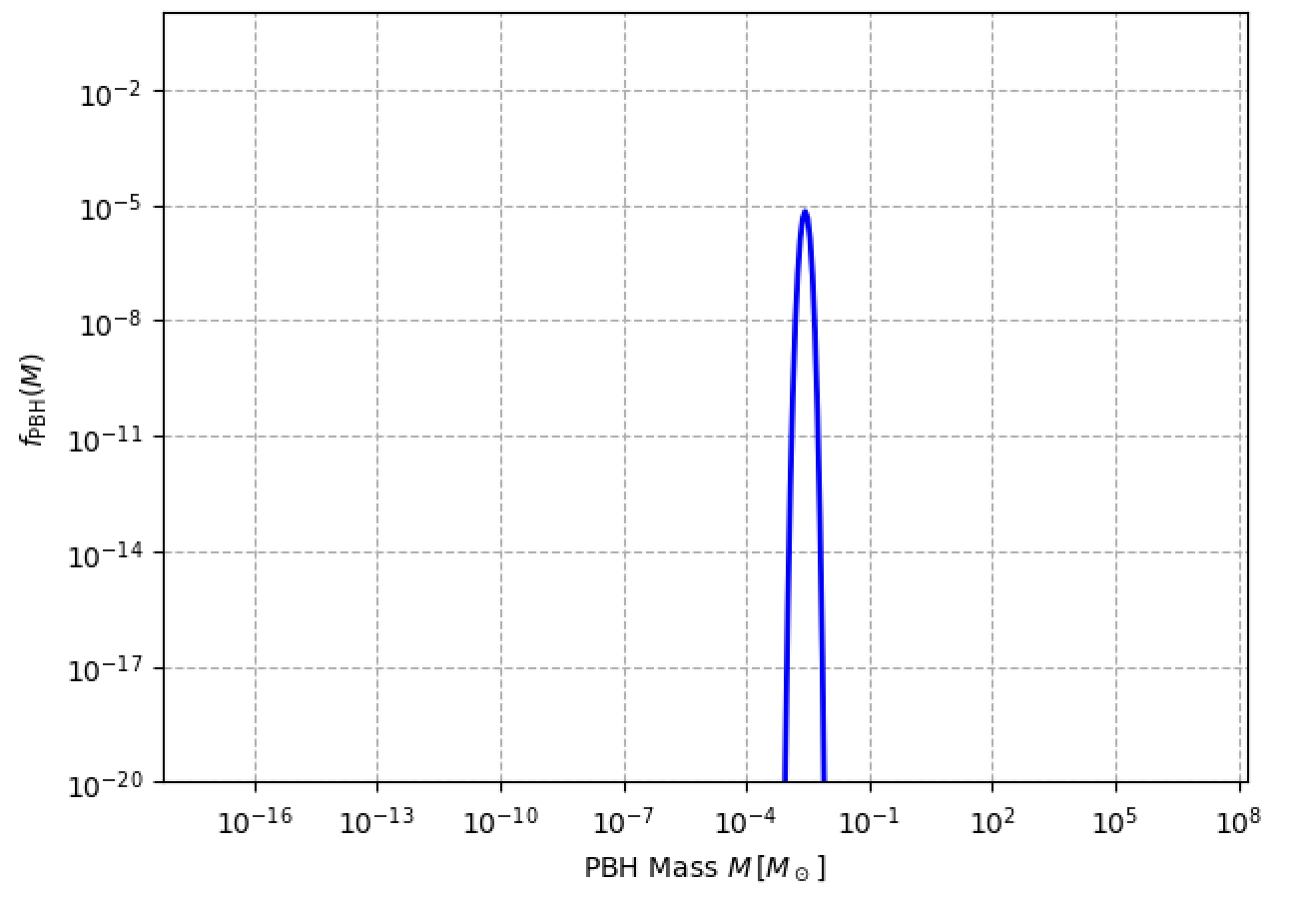}
        \subcaption{Monochromatic dark matter fraction $f_{\mathrm{PBH}}^{\mathrm{mono}}(M)$.}
        \label{fig:fpbh_mass}
    \end{minipage}
    \vspace{0.4cm}
    \begin{minipage}[b]{0.6\textwidth}
        \centering
        \includegraphics[width=\textwidth]{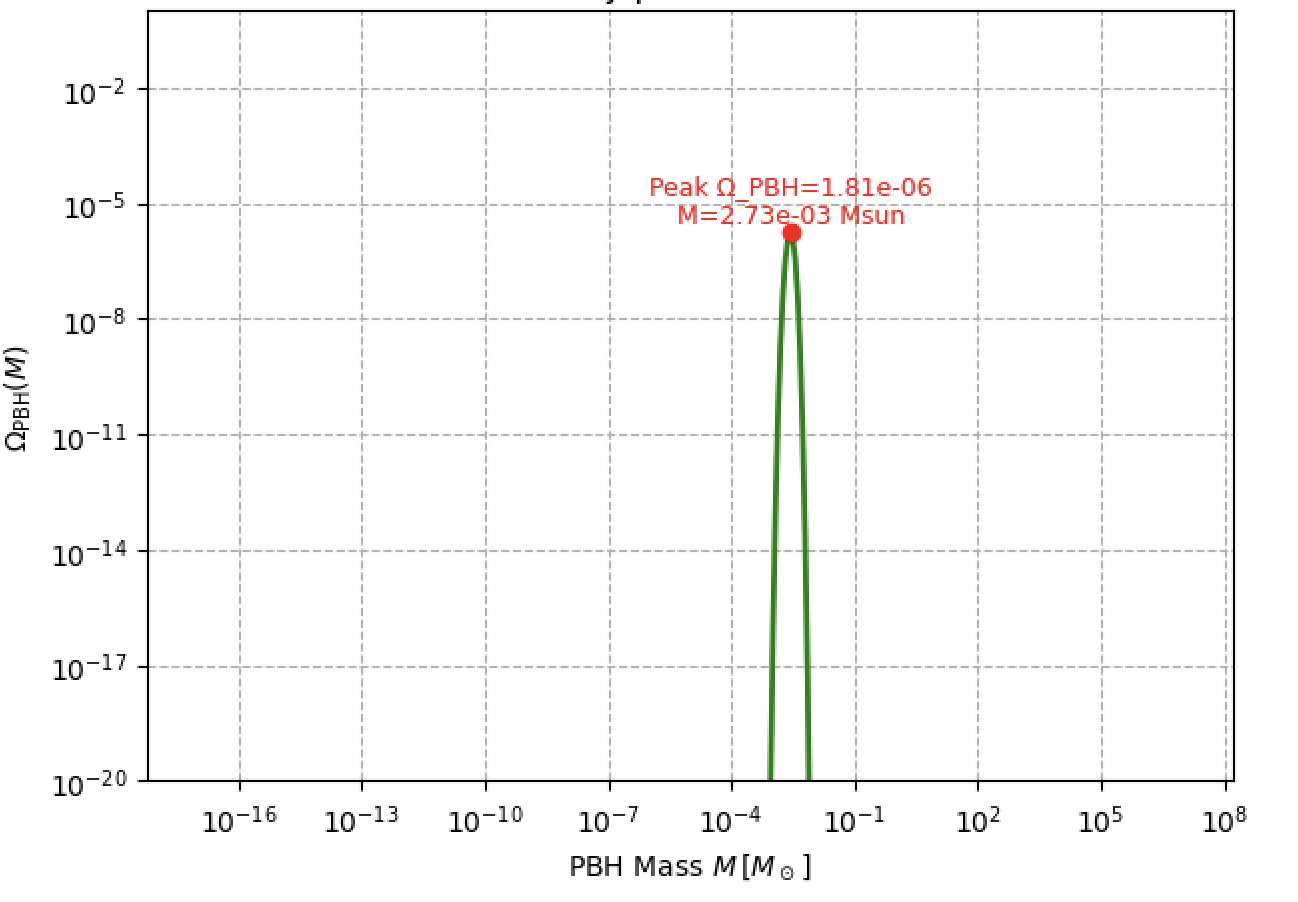}
        \subcaption{Density parameter $\Omega_{\mathrm{PBH}}^{\mathrm{mono}}(M)$.}
        \label{fig:omega_mass}
    \end{minipage}
    \caption{Monochromatic PBH production functions evaluated at horizon entry for baseline parameters $q = 0.237$, $N = 33.74$, and $H_{\mathrm{end}} = 1493.901~\mathrm{GeV}$: (a) initial mass fraction $\beta(M)$, (b) present-day monochromatic energy fraction $f_{\mathrm{PBH}}^{\mathrm{mono}}(M)$ peaking at $M = 2.73 \times 10^{-3}\,M_\odot$ with $f_{\mathrm{PBH}}^{\mathrm{mono}} \approx 6.83 \times 10^{-6}$, and (c) density parameter $\Omega_{\mathrm{PBH}}^{\mathrm{mono}}(M)$ peaking at $\Omega_{\mathrm{PBH}} \approx 1.81 \times 10^{-6}$.}
    \label{fig:monochromatic_trio}
\end{figure}

\subsection{Formalism II: Critical Collapse and Continuous Differential Spectrum}
\label{subsec:critical_collapse_formalism}

Spherical hydrodynamic simulations show that the final mass $M$ of a PBH formed from an overdensity $\delta$ exceeding the critical threshold $\delta_c \equiv \delta_B(k)$ obeys universal near-threshold scaling\cite{PhysRevD.94.084012,PhysRevLett.70.9}:
\begin{equation}
M = K M_H(T_f) \left( \delta - \delta_c \right)^{\gamma_{\mathrm{crit}}},
\label{eq:critical_scaling}
\end{equation}
where $K \approx 4.0$ is a dimensionless scaling prefactor and $\gamma_{\mathrm{crit}} \approx 0.356$ is the universal critical exponent during radiation domination\cite{Musco_2009}. We explicitly distinguish the critical scaling exponent $\gamma_{\mathrm{crit}}$ from the monochromatic collapse efficiency $\gamma_c$ defined in Eq.~\eqref{eq:monochromatic_mass_relation}.

Because near-threshold scaling generates a distribution of black hole masses $M \le M_H$ from a single horizon re-entry scale, the initial collapse fraction per scale $R$ is distributed continuously over mass\cite{Byrnes_2018}. Defining the reduced overdensity variable $x \equiv \delta - \delta_c = (M / K M_H)^{1/\gamma_{\mathrm{crit}}}$, the Jacobian transforming overdensity space into logarithmic mass space yields $d\delta / d\ln M = x / \gamma_{\mathrm{crit}}$. The initial differential mass fraction $d\beta(R) / d\ln M$ produced at formation by a single parent scale $R$ is\cite{Byrnes_2018}
\begin{equation}
\frac{d\beta(R)}{d\ln M} = \left( \frac{M}{M_H} \right) P(\delta) \frac{d\delta}{d\ln M} = \frac{1}{\sqrt{2\pi}\sigma_B(R)} \left( \frac{M}{M_H} \right) \left( \frac{x}{\gamma_{\mathrm{crit}}} \right) \exp\left[ -\frac{(\delta_c + x)^2}{2\sigma_B^2(R)} \right].
\label{eq:dBeta_dlnM}
\end{equation}

Applying radiation redshifting yields the present-day differential mass fraction $df_{\mathrm{PBH}}(R) / d\ln M$ sourced by parent scale $R$\cite{Green_2021}:
\begin{equation}
\frac{df_{\mathrm{PBH}}(R)}{d\ln M} = \left( \frac{1}{3.27 \times 10^{-8}} \right) \left( \frac{106.75}{g_{*,f}(R)} \right)^{1/4} \left( \frac{M}{M_\odot} \right)^{-1/2} \frac{d\beta(R)}{d\ln M}.
\label{eq:dfPBH_per_R}
\end{equation}

Because an extended power spectrum induces critical collapse across a continuous range of horizon entry times, mode scales $R = k^{-1}$ re-enter at distinct temperatures $T_f(R)$. Summing these independent contributions across all active parent smoothing scales $\ln R$ yields the total present-day differential abundance spectrum $df_{\mathrm{PBH}} / d\ln M$\cite{Byrnes_2018,Green_2021}:
\begin{equation}
\frac{df_{\mathrm{PBH}}}{d\ln M} = \int d(\ln R) \frac{df_{\mathrm{PBH}}(R)}{d\ln M}.
\label{eq:total_spectrum_integral}
\end{equation}
The total integrated PBH dark matter fraction $f_{\mathrm{PBH}}^{\mathrm{tot}}$ across the entire mass distribution is obtained via
\begin{equation}
f_{\mathrm{PBH}}^{\mathrm{tot}} = \int d(\ln M) \frac{df_{\mathrm{PBH}}}{d\ln M}.
\label{eq:f_tot_integral}
\end{equation}

\begin{figure}[htbp]
    \centering
    \includegraphics[width=0.8\textwidth]{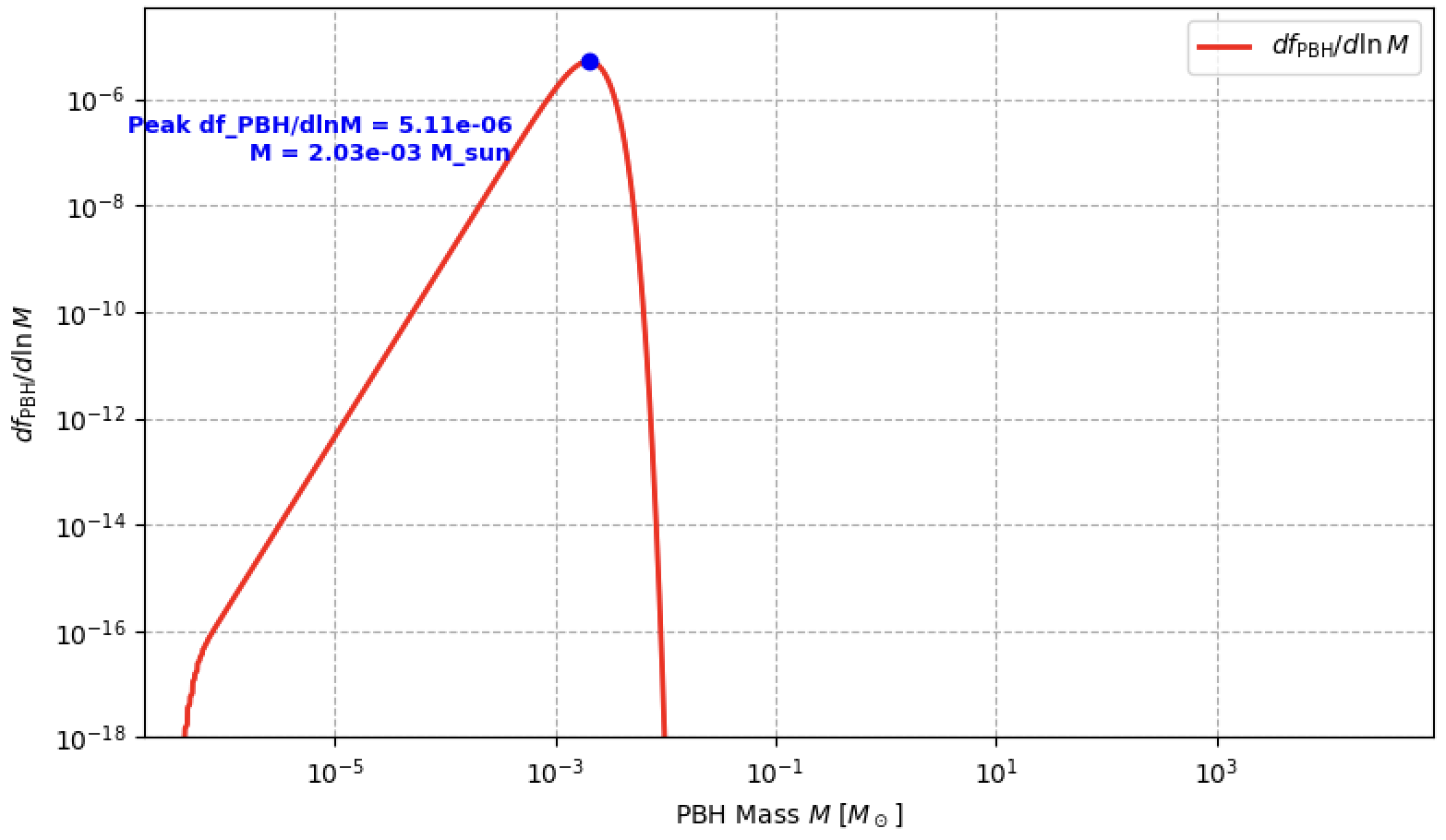}
    \caption{Total integrated differential mass spectrum $df_{\mathrm{PBH}}/d\ln M$ obtained via critical collapse integration over active parent scales $\ln R$ for baseline parameters $q = 0.237$, $N = 33.74$, and $H_{\mathrm{end}} = 1493.901~\mathrm{GeV}$. Near-threshold scaling produces an asymmetric low-mass power-law tail ($M \ll M_H$), reaching a peak value of $df_{\mathrm{PBH}}/d\ln M = 5.11 \times 10^{-6}$ at $M = 2.03 \times 10^{-3}\,M_\odot$. Integrating under this differential distribution yields a total integrated PBH dark matter fraction of $f_{\mathrm{PBH}}^{\mathrm{tot}} \approx 1.02 \times 10^{-5}$.}
    \label{fig:critical_collapse_spectrum}
\end{figure}

\subsection{Physical Comparison and Formalism Limitations}
\label{subsec:formalism_comparison}

The three abundance metrics and two underlying formalisms provide distinct physical roles:
\begin{itemize}
    \item \textbf{Initial vs.\ Present-Day Metrics:} The initial collapse fraction $\beta(R)$ measures the local collapse probability at horizon entry ($\eta = \eta_f$). In contrast, $f_{\mathrm{PBH}}^{\mathrm{mono}}(M)$ and $d f_{\mathrm{PBH}} / d\ln M$ incorporate $a(T_f)$ radiation redshifting to compute present-day dark matter fractions\cite{PhysRevD.81.104019,Green_2021}.
    
    \item \textbf{Peak Scale Identification:} The monochromatic approximation $f_{\mathrm{PBH}}^{\mathrm{mono}}(M)$ assumes a delta-function scale-to-mass mapping $M = \gamma_c M_H$. This prescription serves as a computationally efficient proxy for parameter scans. It reliably identifies the characteristic peak mass of PBH production \cite{Sasaki_2018}.
    
    \item \textbf{Spectral Shape and Observational Constraints:} The critical-collapse mass function (Formalism~II) accounts for near-threshold scaling ($\delta \to \delta_c^+$). This mechanism generates a low-mass power-law tail ($M \ll M_H$) for each parent scale $R$\cite{Byrnes_2018}. Consequently, the total differential spectrum $df_{\mathrm{PBH}}/d\ln M$ in Eq.~\eqref{eq:total_spectrum_integral} provides the physical distribution. This spectrum is required for direct comparisons against observational microlensing, Hawking evaporation, and gravitational-wave exclusion limits \cite{Carr_2021, Green_2021}.
    
    \item \textbf{Comparison of Monochromatic vs.\ Critical-Collapse Peak Values:} We emphasize that the numerical values reported in Figure~\ref{fig:monochromatic_trio} and Figure~\ref{fig:critical_collapse_spectrum} describe distinct physical quantities and are not directly interchangeable:
    \begin{enumerate}
        \item \emph{Peak Mass Shift:} Figure~\ref{fig:monochromatic_trio} reports a monochromatic peak at $M = 2.73 \times 10^{-3}\,M_\odot$, corresponding to a fixed horizon mass fraction $M = \gamma_c M_H$. In contrast, Figure~\ref{fig:critical_collapse_spectrum} reports a critical-collapse peak at $M = 2.03 \times 10^{-3}\,M_\odot$. This shift toward lighter mass scales occurs because near-threshold collapse ($\delta \to \delta_c^+$) obeys $(\delta - \delta_c)^{\gamma_{\mathrm{crit}}} < 1$. This scaling systematically suppresses black hole masses relative to $M_H$ and populates a characteristic low-mass power-law tail ($M \propto (\delta - \delta_c)^{\gamma_{\mathrm{crit}}}$)\cite{PhysRevD.58.107502,Niemeyer_1998}.
        
        \item \emph{Differential Peak Height vs.\ Total Integrated Fraction:} Figure~\ref{fig:monochromatic_trio}(b) reports a monochromatic dark matter fraction peak of $f_{\mathrm{PBH}}^{\mathrm{mono}} \approx 6.83 \times 10^{-6}$ (or $\Omega_{\mathrm{PBH}} \approx 1.81 \times 10^{-6}$). This value assumes a single fixed mass per scale. Conversely, Figure~\ref{fig:critical_collapse_spectrum} gives a peak differential spectral amplitude of $df_{\mathrm{PBH}}/d\ln M = 5.11 \times 10^{-6}$. This metric measures the dark matter fraction per logarithmic mass interval.

        We perform the logarithmic integration in Eq.~\eqref{eq:f_tot_integral} across this continuous distribution:
        \begin{equation}
        f_{\mathrm{PBH}}^{\mathrm{tot}} = \int \frac{df_{\mathrm{PBH}}}{d\ln M} d(\ln M) \approx 1.02 \times 10^{-5}.
        \label{eq:f_tot_evaluated}
        \end{equation}
        This result corresponds to a total integrated density parameter $\Omega_{\mathrm{PBH}}^{\mathrm{tot}} = f_{\mathrm{PBH}}^{\mathrm{tot}} \Omega_{\mathrm{DM},0} \approx 2.70 \times 10^{-6}$. The differential peak spectral height ($5.11 \times 10^{-6}$) is lower than the monochromatic proxy ($6.83 \times 10^{-6}$). However, integrating across the continuous power-law tail increases the total dark matter fraction to $f_{\mathrm{PBH}}^{\mathrm{tot}} \approx 1.02 \times 10^{-5}$. This outcome confirms that the baseline model yields a subdominant dark matter contribution ($f_{\mathrm{PBH}}^{\mathrm{tot}} \ll 1$).
    \end{enumerate}
\end{itemize}
%%%%%%%%%%%%%%%%%%%%%%%%%%%%%%%%%%%%%%%%%%%%%%%%%%%%%%%%%%%%
\section{Observational Constraints and Benchmarks Predictions}
\label{sec:observational_constraints}

To assess the physical viability of PBHs generated via vector-field magnetogenesis during the non-singular bouncing cosmology, we compare our predicted differential abundance $f_{\mathrm{PBH}}(M)$ against current observational exclusions across the mass range $M \in [10^{-18}, 10^8]\,M_\odot$ \cite{Carr_2021,Green_2021}. Below, we present our benchmark spectra, compare them with observational bounds, and analyze the joint parameter dependencies that distinguish subdominant PBH populations from scenarios that can account for all dark matter \cite{Carr_2021, Green_2021}.

% ==========================================================
\subsection{Final PBH Mass Spectrum and Benchmark Parameter Set}
\label{subsec:benchmark_spectrum}

The present-day energy fraction $f_{\mathrm{PBH}}(M) \equiv (1/\Omega_{\mathrm{DM}}) (d\Omega_{\mathrm{PBH}}/d\ln M)$ is evaluated from the horizon-crossing curvature spectrum $\mathcal{P}_\zeta^{\mathrm{HC}}(k)$ filtered through the scale-dependent collapse threshold $\delta_B(k)$\cite{Kushwaha_2024}. For our primary benchmark configuration, we assume coupling parameters $q = 0.237$, spectral exponent $n = 10$, e-folding duration $N = 33.74$, and end-of-contraction scale $H_{\mathrm{end}} = 1493.901~\mathrm{GeV}$\cite{PhysRevD.94.083524}.

Evaluating the critical-collapse differential spectrum over the active momentum band $k \in [10^{-39}, 10^{-30}]\,\mathrm{GeV}$ yields a localized abundance peak centered at a sub-solar mass $M \approx 2.7 \times 10^{-3}\,M_\odot$. The maximum energy fraction reaches $f_{\mathrm{PBH}} \approx 10^{-5}$. This benchmark represents a predicted subdominant PBH population, where black holes constitute approximately $0.001\%$ of the total cold dark matter density ($f_{\mathrm{PBH}}^{\mathrm{tot}} \ll 1$)\cite{Carr_2021}. Such sub-solar black holes avoid overclosing the Universe while remaining abundant enough to be relevant for future gravitational-lensing or binary-capture searches\cite{Niikura_2019,Clesse_2017}.

% ==========================================================
\subsection{Comparison with Observational Constraints}
\label{subsec:obs_comparison}

PBH abundances are tightly bound by distinct cosmological and astrophysical observations depending on the mass scale $M$\cite{Carr_2021,Green_2021,PBHbounds}:

\begin{itemize}
    \item \textbf{Hawking Evaporation Limits ($M \lesssim 10^{-16}\,M_\odot$):} PBHs with initial masses $M \lesssim 10^{-17}\,M_\odot$ evaporate within a Hubble time. Bounds in this regime are established by isotropic extragalactic gamma-ray background (EGRB) measurements \cite{PhysRevD.81.104019}, Galactic Center 511\,keV electron-positron annihilation lines \cite{PhysRevLett.123.251102,PhysRevLett.123.251101}, Voyager~1 cosmic-ray $e^\pm$ fluxes \cite{PhysRevLett.122.041104}, Galactic Center X-ray emissions\cite{Inoue:2017_Xray,Manshanden:2018_MW_Searches}, and 21-cm neutral hydrogen absorption signals during Cosmic Dawn (EDGES) \cite{PhysRevD.98.043006}.
    
    \item \textbf{Gravitational Microlensing ($10^{-11}\,M_\odot \lesssim M \lesssim 10^1\,M_\odot$):} Transient photometric magnification of background stars restricts the compact-object fraction of galactic halos\cite{Alcock_2000}. Upper bounds are enforced by high-cadence Subaru Hyper Suprime-Cam (HSC) M31 observations \cite{Niikura_2019}, Kepler optical light curves \cite{Griest_2014}, EROS/MACHO survey constraints \cite{Alcock_2000,Tisserand_2007}, Icarus stellar magnification events\cite{Oguri:2017ock}, and Optical Gravitational Lensing Experiment (OGLE) surveys \cite{Niikura_2019}.
    
    \item \textbf{Gravitational Wave and Dynamical Bounds ($M \gtrsim 10^0\,M_\odot$):} At intermediate and solar mass scales, constraints arise from LIGO-Virgo-KAGRA binary merger rates \cite{PhysRevLett.117.061101,PhysRevLett.123.161102,Clesse_2017}, stochastic gravitational-wave background limits\cite{Wang:2018_LIGO_SGWB_PBH}, wide-binary star disruption \cite{Monroy_Rodr_guez_2014}, and dwarf galaxy dynamical heating\cite{Koushiappas:2017_DwarfDynamics}.
\end{itemize}

\begin{figure}[htbp]
    \centering
    \includegraphics[width=0.8\textwidth]{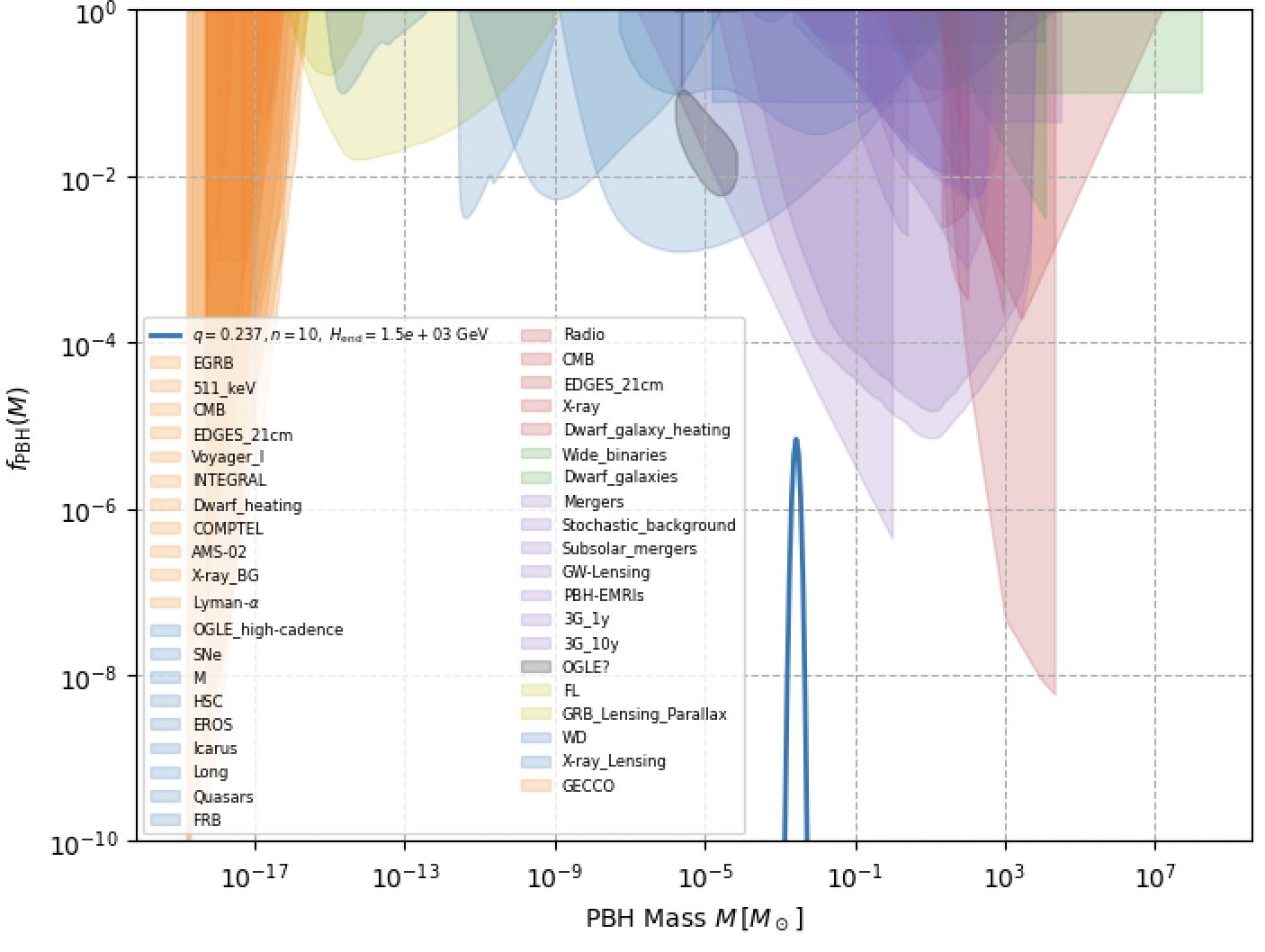}
    \caption{Sub-solar benchmark spectrum ($q=0.237, n=10, H_{\mathrm{end}} \approx 1.5\times 10^3\,\mathrm{GeV}$ at the end of contraction) peaking at $M \approx 2.7 \times 10^{-3}\,M_\odot$ with $f_{\mathrm{PBH}} \approx 10^{-5}$. The prediction is contrasted against grouped observational exclusion boundaries: Hawking evaporation limits (EGRB, 511\,keV, Voyager-1), gravitational microlensing constraints (Subaru HSC, Kepler, EROS, OGLE), and dynamical/GW bounds (LIGO-Virgo-KAGRA, wide binaries) \cite{PhysRevLett.107.231101,Niikura_2019,PBHbounds}. While observationally consistent (lying more than two orders of magnitude below microlensing limits), this benchmark forms a subdominant dark matter component\cite{Green_2021}.}
    \label{fig:fpbh_obs_single}
\end{figure}

As shown in Figure~\ref{fig:fpbh_obs_single}, our sub-solar benchmark ($M \approx 2.7 \times 10^{-3}\,M_\odot$, $f_{\mathrm{PBH}} \approx 10^{-5}$) sits at a mass scale where combined microlensing limits (from Kepler, EROS, and OGLE) restrict compact dark matter fractions to $f_{\mathrm{PBH}} \lesssim 10^{-3} - 10^{-2}$\cite{Tisserand_2007,Griest_2014,Niikura_2019}. Our peak abundance of $f_{\mathrm{PBH}} \approx 10^{-5}$ lies more than two orders of magnitude below these observational exclusion boundaries, confirming that the model satisfies all observational bounds\cite{Carr_2021}. We emphasize that lying safely below observational bounds establishes physical consistency and rules out overclosure, but does not automatically imply that the model constitutes the dominant dark matter candidate; for this benchmark, the resulting population is strictly subdominant ($f_{\mathrm{PBH}}^{\mathrm{tot}} \approx 10^{-5} \ll 1$).

\begin{figure}[htbp]
    \centering
    \includegraphics[width=0.8\textwidth]{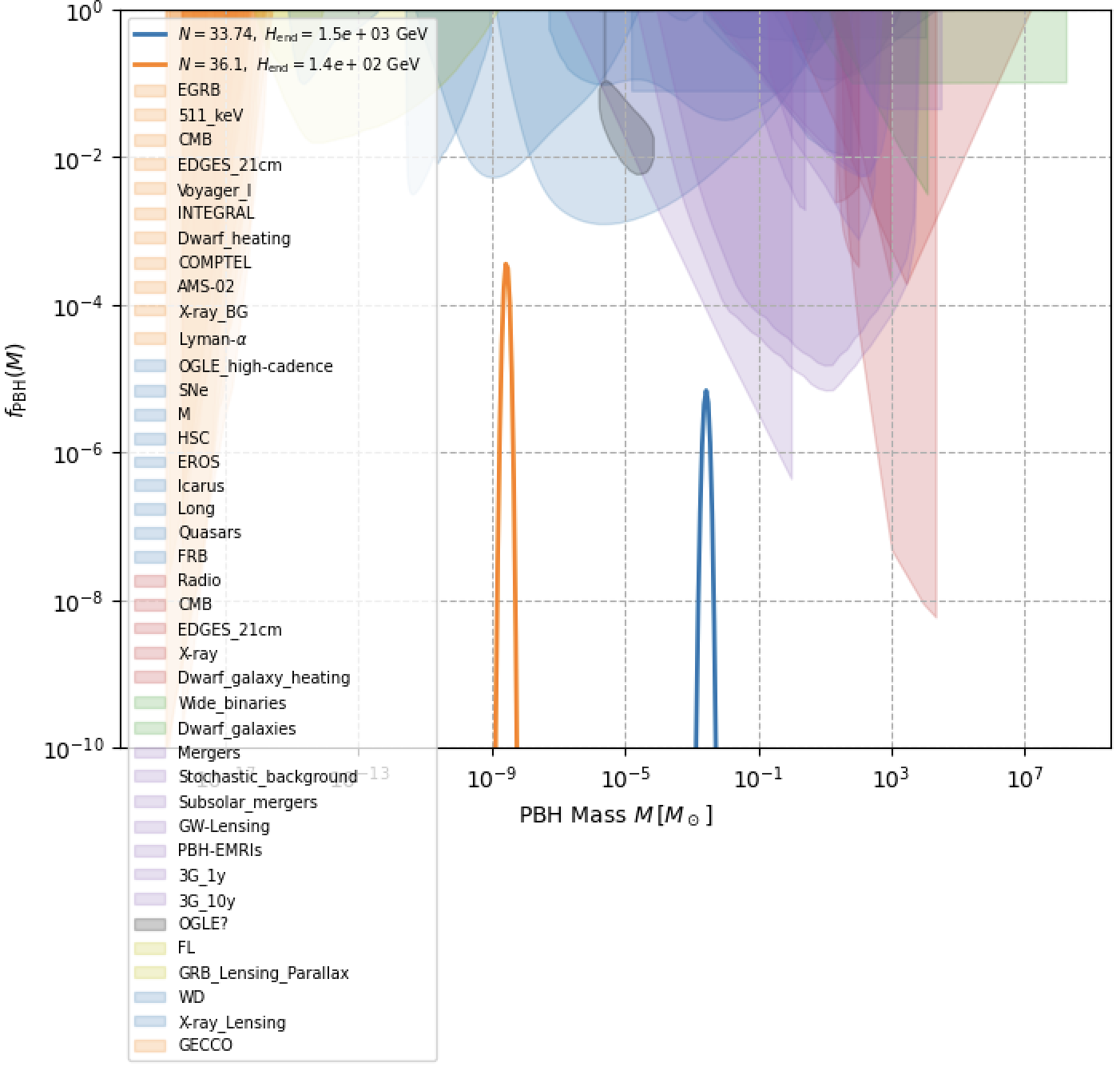}
    \caption{Displacement of the PBH mass spectrum from the sub-solar planetary window ($M \sim 10^{-3}\,M_\odot, N=33.74$) to the sub-planetary scale ($M \sim 3 \times 10^{-9}\,M_\odot, N=36.10$), achieved by adjusting the contracting duration $N$ (modifying $H_{\mathrm{end}}$ at the end of contraction) and extending $k_{\max}$ for modes re-entering during radiation domination. Observational constraints are grouped into simplified background shaded regions to maintain visual legibility across parameter comparisons\cite{Green_2021,PBHbounds}.}
    \label{fig:fpbh_obs_two}
\end{figure}
% ==========================================================
\subsection{Parameter Constraints and Overclosure Limits}
\label{subsec:parameter_constraints}

The peak amplitude of the present-day PBH abundance $f_{\mathrm{PBH}}$ exhibits an extreme exponential sensitivity to the underlying background parameters of the non-singular bouncing cosmology, specifically the Hubble scale at the end of contraction $H_{\mathrm{end}}$ and the contraction parameter $q$ \cite{PhysRevD.94.083524, Kushwaha_2024}. This exponential dependence originates directly from the Press--Schechter Gaussian tail integral \cite{Green:2004wb}, where the initial collapse fraction scales as $\beta \propto \exp(-\delta_B^2 / 2\sigma_B^2)$ \cite{Harada:2013epa, Young:2014ana}. Because the smoothed density variance $\sigma_B(R)$ scales linearly with the curvature perturbation amplitude $\mathcal{P}_\zeta^{1/2}(k)$, modest variations in the background parameters that alter $\mathcal{P}_\zeta(k)$ translate into vast, multi-order-of-magnitude shifts in $f_{\mathrm{PBH}}$ \cite{Sasaki_2018,Green_2021}.

\begin{figure}[htbp]
    \centering
    \begin{minipage}[b]{0.9\textwidth}
        \centering
        \includegraphics[width=\textwidth]{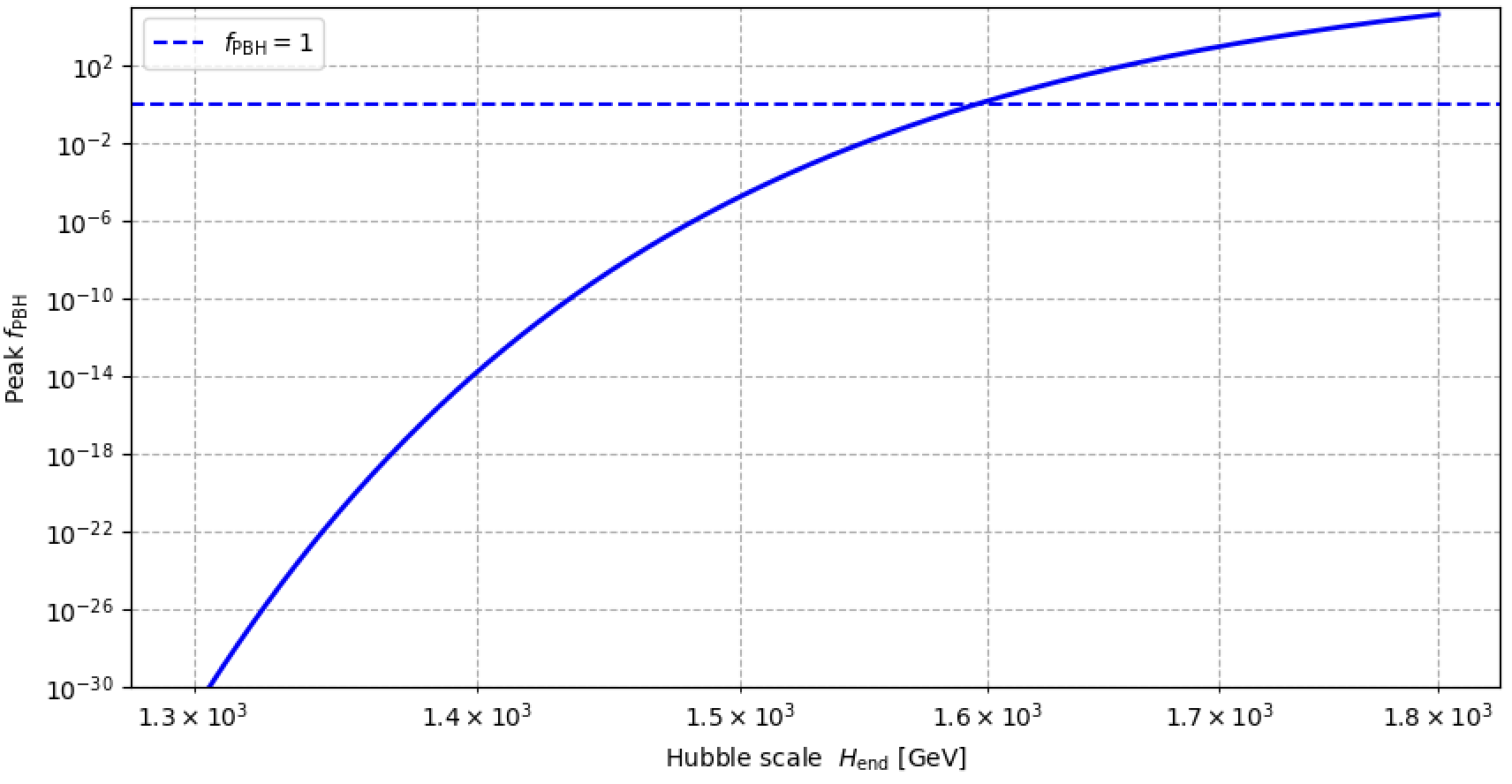}
        \subcaption{Peak $f_{\mathrm{PBH}}$ vs.\ Hubble scale $H_{\mathrm{end}}$.}
        \label{fig:peak_fpbh_vs_Hend}
    \end{minipage}
    \hfill
    \begin{minipage}[b]{0.7\textwidth}
        \centering
        \includegraphics[width=\textwidth]{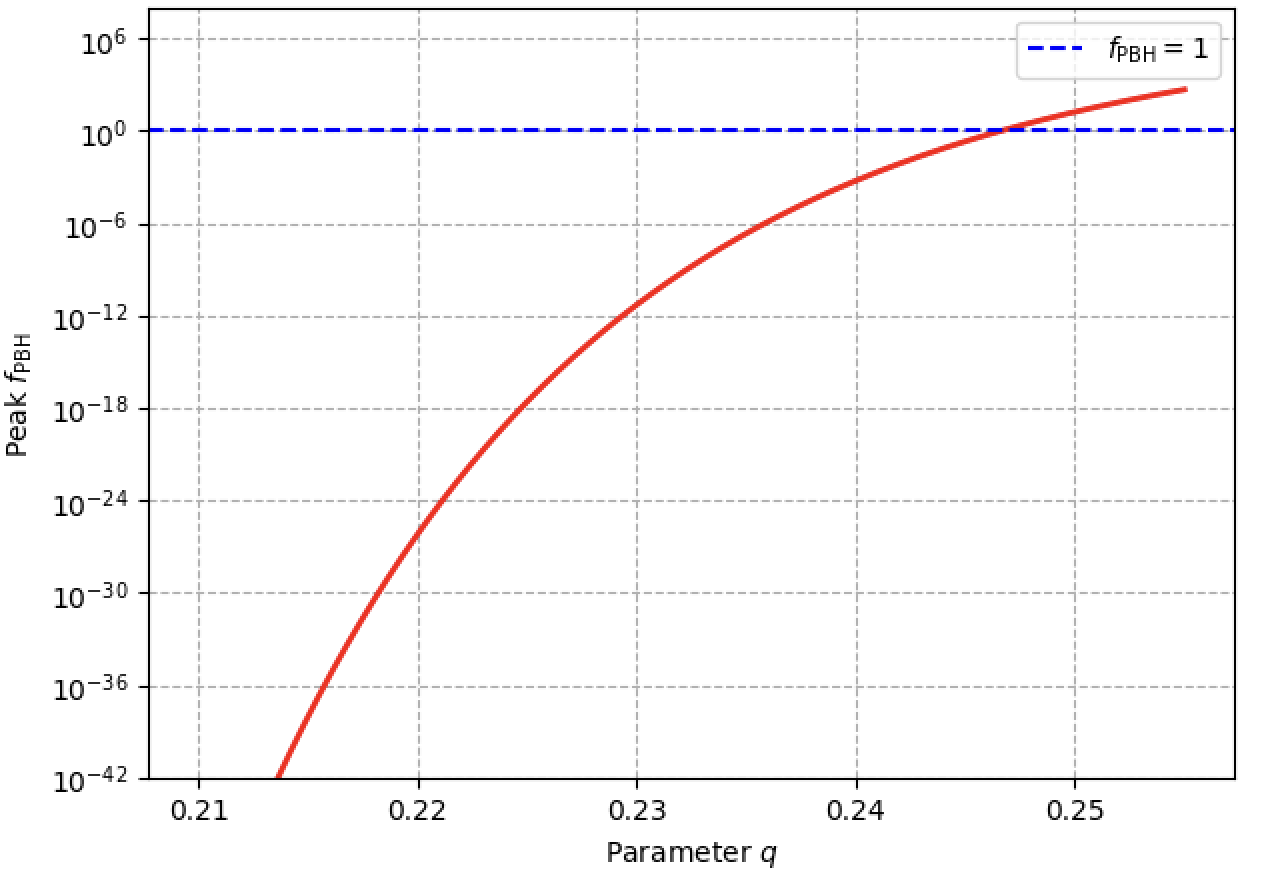}
        \subcaption{Peak $f_{\mathrm{PBH}}$ vs.\ contraction parameter $q$.}
        \label{fig:peak_fpbh_vs_q}
    \end{minipage}
    \caption{Dependence of the peak PBH dark matter fraction $f_{\mathrm{PBH}}$ on the primary background parameters of the ekpyrotic contracting phase: (a) variation with the end-of-contraction Hubble scale $H_{\mathrm{end}}$ (blue solid line), crossing the total dark matter threshold $f_{\mathrm{PBH}} = 1$ (horizontal dashed line) near $H_{\mathrm{end}} \approx 1.6 \times 10^3\,\mathrm{GeV}$; (b) variation with the contraction parameter $q$ (red solid line), demonstrating steep exponential growth and saturating $f_{\mathrm{PBH}} = 1$ at $q \approx 0.247$. Parameter choices yielding $f_{\mathrm{PBH}} > 1$ are excluded by cosmological dark matter overclosure constraints \cite{Carr_2021,Green_2021}.}
    \label{fig:parameter_constraints_dual}
\end{figure}

Figure~\ref{fig:parameter_constraints_dual}(a) illustrates the response of peak $f_{\mathrm{PBH}}$ to variations in the Hubble scale $H_{\mathrm{end}}$ at the end of contraction. As $H_{\mathrm{end}}$ increases from $1.3 \times 10^3\,\mathrm{GeV}$ to $1.8 \times 10^3\,\mathrm{GeV}$. The horizontal dashed line marks the physical saturation boundary $f_{\mathrm{PBH}} = 1$, corresponding to the scenario where PBHs could potentially account for the total dark
matter \cite{Carr_2021,Green_2021}. Requiring that PBHs do not overclose the Universe ($f_{\mathrm{PBH}} \le 1$) imposes a strict physical upper bound on the Hubble scale at the end of contraction \cite{PhysRevD.94.083524}:
\begin{equation}
H_{\mathrm{end}} \le H_{\mathrm{end}}^{\mathrm{max}} \approx 1.6 \times 10^3\,\mathrm{GeV},
\label{eq:Hend_overclosure_bound}
\end{equation}
which is consistent with the electromagnetic backreaction saturation limit ($H_{\mathrm{end}} \le 1493.901~\mathrm{GeV}$) enforced to maintain the background ekpyrotic contraction \cite{Demozzi_2009, PhysRevD.94.083524, Kushwaha_2024}.

Similarly, Figure~\ref{fig:parameter_constraints_dual}(b) depicts the dependence on the contraction parameter $q$. The index $q$ governs the scale-factor evolution $a(\eta) \propto (\tilde{\eta}_{\mathrm{end}} - \eta)^{\tilde{q}}$ during the ekpyrotic phase, directly setting the magnetic spectral index $n_B$ and the amplitude prefactor $C_B$ \cite{PhysRevD.94.083524}. A small increase in $q$ from $0.21$ to $0.255$ causes peak $f_{\mathrm{PBH}}$ to surge from $10^{-42}$ to over $10^3$. The dark matter overclosure condition $f_{\mathrm{PBH}} = 1$ places a tight upper limit on the contraction exponent:
\begin{equation}
q \le q_{\mathrm{max}} \approx 0.247.
\label{eq:q_overclosure_bound}
\end{equation}

Together, Equations~\eqref{eq:Hend_overclosure_bound} and \eqref{eq:q_overclosure_bound} delineate the viable parameter space wherein vector-field magnetogenesis in a non-singular bouncing cosmology generates physical, non-overclosing PBH populations or accounts for the total dark matter abundance \cite{Carr_2021, Green_2021}.

\subsection{Joint Parameter Sensitivity: Wavenumber Cutoffs, Contraction Duration, and Dark Matter Scenarios}
\label{subsec:parameter_dependence_windows}

The background parameters ($N, H_{\mathrm{end}}$) at the end of contraction and the wavenumber cutoff $[k_{\min}, k_{\max}]$ jointly govern $f_{\mathrm{PBH}}(M)$. Together, they determine its mass location and peak amplitude \cite{PhysRevD.94.083524,Kushwaha_2024}. During radiation domination, the horizon-mass relation scales as $M_H \propto T_f^{-2} \propto k^{-2}$. This relation directly maps the active wavenumber range to the black hole mass position. Consequently, shifting the mass spectrum to lighter scales requires expanding the upper wavenumber cutoff $k_{\max}$. This expansion incorporates higher-frequency modes that re-enter earlier in the radiation-dominated era. Simultaneously, the contracting duration $N$ determines the Hubble scale at the end of contraction ($H_{\mathrm{end}} \propto e^{-N}$). It also controls the curvature prefactor ($C_C \propto e^{-8.42 N}$), which scales the peak abundance $f_{\mathrm{PBH}}$.

Figure~\ref{fig:fpbh_obs_single} and Figure~\ref{fig:fpbh_obs_two} illustrate how these parameter variations relocate the abundance peak. Varying $N$ and the wavenumber cutoff $[k_{\min}, k_{\max}]$ shifts the peak across distinct mass regimes:

\begin{itemize}
    \item \textbf{Sub-solar Benchmark ($N = 33.74$, $k \in [10^{-39}, 10^{-30}]\,\mathrm{GeV}$):} At the end of contraction, the Hubble scale evaluates to $H_{\mathrm{end}} \approx 1.5 \times 10^3\,\mathrm{GeV}$. This configuration produces a peak centered in the planet-mass regime at $M \approx 2.7 \times 10^{-3}\,M_\odot$ with an abundance of $f_{\mathrm{PBH}} \approx 10^{-5}$.
    
    \item \textbf{Sub-planetary Benchmark ($N = 36.10$, $k \in [10^{-39}, 10^{-27}]\,\mathrm{GeV}$):} Extending the upper cutoff to $k_{\max} = 10^{-27}\,\mathrm{GeV}$ incorporates higher-frequency modes that re-enter earlier during radiation domination. This shift lowers the mass peak to the sub-planetary scale $M \approx 3 \times 10^{-9}\,M_\odot$. Simultaneously, increasing $N$ to $36.10$ reduces $H_{\mathrm{end}}$ to $1.4 \times 10^2\,\mathrm{GeV}$ at the end of contraction. It enhances the peak abundance to $f_{\mathrm{PBH}} \approx 3.5 \times 10^{-4}$, remaining safely below Subaru HSC microlensing bounds ($f_{\mathrm{PBH}} \lesssim 10^{-3}$) \cite{Niikura_2019}.
\end{itemize}

We evaluate observational consistency and dark matter contributions across two physical regimes. These scenarios are obtained by adjusting the wavenumber cutoff and the contracting duration $N$:

\begin{enumerate}
    \item \textbf{Subdominant PBH Population ($f_{\mathrm{PBH}} \ll 1$):} We assume $N \approx 33.7\text{--}36.1$ and $k_{\max} \sim 10^{-30}\text{--}10^{-27}\,\mathrm{GeV}$. This configuration generates planet-mass ($M \sim 10^{-3}\,M_\odot$) or sub-planetary ($M \sim 10^{-9}\,M_\odot$) black holes. These benchmarks lie $10^1\text{--}10^2$ times below microlensing limits and are fully consistent with observations. However, they form a subdominant fraction of cold dark matter ($f_{\mathrm{PBH}} \sim 10^{-5}\text{--}10^{-4}$)\cite{Carr_2021,Green_2021}.
    
    \item \textbf{Full Dark Matter Scenario ($f_{\mathrm{PBH}} \sim 1$):} PBHs can account for the entirety of cold dark matter ($f_{\mathrm{PBH}}^{\mathrm{tot}} = 1$) if two conditions are met simultaneously. First, the wavenumber cutoff must shift further into the UV ($k_{\max} \sim 10^{-24}\text{--}10^{-21}\,\mathrm{GeV}$). This places the mass peak inside an asteroid-mass window ($10^{-16}\,M_\odot \lesssim M \lesssim 10^{-11}\,M_\odot$) during radiation domination\cite{Carr_2021}. Second, the contracting duration $N$ must be adjusted to scale the peak abundance to $f_{\mathrm{PBH}} = 1$ \cite{Carr_2021}. Under these conditions, the scenario satisfies all observational constraints and could potentially account for the total dark matter
abundance, subject to further analysis\cite{Carr_2021,Green_2021}.
\end{enumerate}
% ==========================================================
% FIGURE ENVIRONMENTS FOR OBSERVATIONAL PLOTS (STACKED FORMAT FOR CLARITY)
% ==========================================================

%%%%%%%%%%%%%%%%%%%%%%%%%%%%%%%%%%%%%%%%%%%%%%%%%%%%%%%%%%%%
\section{Conclusions}
\label{sec:conclusions}
In this work, we investigated the generation of primordial black holes (PBHs) from primordial magnetic fields (PMFs) produced during the contracting phase of a non-singular bouncing cosmology \cite{PhysRevD.94.083524,Kushwaha_2024}. Non-minimal gauge coupling breaks Maxwellian conformal invariance, amplifying quantum vacuum fluctuations into super-horizon magnetic fields. Upon transitioning into radiation domination, the anisotropic stress of the magnetic field sources magnetic-field-induced curvature perturbations. Our calculation was organized into three distinct stages: matching mode functions continuously across the non-singular bounce\cite{Cai_2012,Membiela:2013cea}, evaluating the momentum convolution to obtain the horizon-crossing curvature power spectrum and smoothed density variance with explicit wavenumber cutoffs\cite{PhysRevD.86.023512,Kushwaha_2024}, and determining the present-day PBH mass distribution using both the monochromatic approximation and the continuous critical-collapse mass function \cite{Green:2004wb, Sasaki_2018}.

Because magnetic anisotropic stress scales quadratically with the vector field, the momentum convolution concentrates power near the upper wavenumber cutoff, establishing a blue-tilted curvature spectrum with an asymptotic $n_\zeta \approx 2.80$ spectral slope \cite{PhysRevD.86.023512,Kushwaha_2024}. This enhanced power drives a peak in the variance at scales re-entering the horizon during the radiation-dominated era \cite{Young:2014ana}. Incorporating near-threshold critical scaling yields a continuous differential mass function characterized by an asymmetric low-mass power-law tail \cite{Niemeyer_1998, PhysRevD.58.107502,Byrnes_2018}. For our reference background scenario, PBH production peaks in the sub-solar mass regime with a total integrated dark matter fraction $f_{\mathrm{PBH}}^{\mathrm{tot}} \ll 1$, lying well below optical microlensing bounds from Subaru HSC and OGLE \cite{Niikura_2019,Green_2021}. Expanding the upper wavenumber cutoff to encompass higher-frequency modes shifts the mass peak into an asteroid-mass window ($10^{-16}\,M_\odot \lesssim M \lesssim 10^{-11}\,M_\odot$), enabling PBHs to potentially account for the total dark matter abundance ($f_{\mathrm{PBH}}^{\mathrm{tot}} = 1$) \cite{Carr_2021}, subject to further observational and statistical analysis. Furthermore, enforcing the dark matter overclosure bound ($f_{\mathrm{PBH}}^{\mathrm{tot}} \le 1$) constrains the primary background parameters to $H_{\mathrm{end}} \le 1.6 \times 10^3~\mathrm{GeV}$ and $q \le 0.247$, establishing a strict physical cutoff consistent with the backreaction saturation threshold \cite{Demozzi_2009, PhysRevD.94.083524, Kushwaha_2024}.

These results demonstrate that vector-field magnetogenesis in a non-singular bouncing cosmology provides a viable, non-inflationary channel for generating physical PBH populations. However, the scope of these conclusions is bounded by specific physical constraints and underlying assumptions. Preserving background ekpyrotic contraction requires the electromagnetic energy density to remain strictly subdominant to the scalar field, placing a firm upper limit on the Hubble scale at the end of contraction \cite{Demozzi_2009, PhysRevD.94.083524}. Additionally, our abundance derivations assume Gaussian statistics for the curvature perturbations, whereas quadratic vector sources inherently generate $\chi^2$-type non-Gaussianities that can reshape the tail of the probability distribution \cite{PhysRevD.86.023512,Young:2014ana}. Incorporating these non-Gaussian statistical corrections\cite{Young:2014ana} and evaluating the induced secondary stochastic gravitational-wave background\cite{Auclair:2022jfa} represent crucial steps to test this mechanism against future observational data from LISA, the Einstein Telescope, and pulsar timing arrays.

% References
% ----------------------------------------------------------------------
\bibliographystyle{JHEP}
\bibliography{references} % Points to your references.bib file

\end{document}